\documentclass[journal]{IEEEtran}
\usepackage{amsmath}
\usepackage{graphicx}
\usepackage{lipsum}

\ifCLASSINFOpdf
\else
\fi
\usepackage{tabularx}
\usepackage{longtable}
\usepackage{cite}
\usepackage{amsmath}
\usepackage{wrapfig}
\usepackage{algorithmic}
\newcolumntype{Y}{>{\centering\arraybackslash}X}
\usepackage{float}
\usepackage{graphics}
\usepackage{graphicx}
\usepackage{caption}
\usepackage{subcaption}
\usepackage{url}
\usepackage{xcolor}
\definecolor{uptxt}{rgb}{0,0,0}
\usepackage{setspace}
\usepackage{tikz}
\usepackage{url}
\usepackage{hyperref}
\def\checkmark{\tikz\fill[scale=0.4](0,.35) -- (.25,0) -- (1,.7) -- (.25,.15) -- cycle;} 
\usepackage{color, colortbl}
\usepackage{multirow}
\usepackage{multicol}
\definecolor{Gr}{gray}{0.8}

\newcommand{\rgbsymbol}{\includegraphics[height=1.5ex]{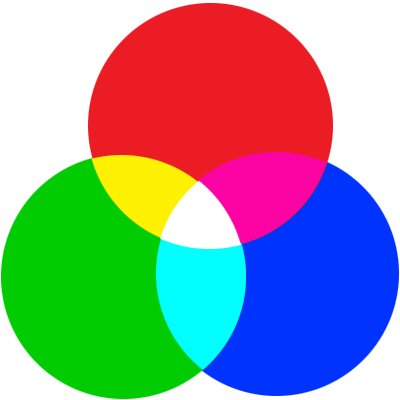}}

\begin{document}
\setlength{\intextsep}{0pt} 
\setlength{\textfloatsep}{6pt} 

\title{Robust Segmentation of Optic Disc and Cup from Fundus Images Using Deep Neural Networks}
%

%

\author{
Aniketh Manjunath, Subramanya Jois, and Chandra Sekhar Seelamantula, \textit{Senior Member, IEEE} 
\thanks{
\newline

\indent{} Manuscript submitted to IEEE Transactions on Image Prcessing on September 29, 2019.  Paper ID: TIP-21464-2019.R1.  This work was supported by the Ministry of Human Resource Development and Indian Council for Medical Research under the IMPRINT India Initiative (Domain: Healthcare; Project No.: 6013). \newline
\indent{} Aniketh Manjunath was with the Department of Electrical Engineering, Indian Institute of Science, Bangalore 560012, India. He is presently with University of Southern California, Los Angeles, CA 90007, USA \textit{(email: anikethm@usc.edu)} \newline
\indent{}Subramanya Jois was with the Department of Electrical Engineering, Indian Institute of Science, Bangalore 560012, India. He is presently with Concordia and McGill Universities (CREATE Innovation program), Montr\'eal, Quebec, Canada (email: sp.subramanya@gmail.com) \newline
\indent{} Chandra Sekhar Seelamantula is with the Department of Electrical Engineering, Indian Institute of Science, Bangalore 560012, India. (Email: css@iisc.ac.in). 
}

}

\markboth{Manuscript Submitted to IEEE Transactions on Image Processing}
\maketitle

\maketitle

\begin{abstract}
Optic disc (OD) and optic cup (OC) are regions of prominent clinical interest in a retinal fundus image. They are the primary indicators of a glaucomatous condition. With the advent and success of deep learning for healthcare research, several approaches have been proposed for the segmentation of important features in retinal fundus images. We propose a novel approach for the simultaneous segmentation of the OD and OC using a {\it residual encoder-decoder network} (REDNet) based regional convolutional neural network (RCNN). The RED-RCNN is motivated by the Mask RCNN (MRCNN). Performance comparisons with the state-of-the-art techniques and extensive validations on standard publicly available fundus image datasets show that RED-RCNN has superior performance compared with MRCNN. RED-RCNN results in Sensitivity, Specificity, Accuracy, Precision, Dice and Jaccard indices of 95.64\%, 99.9\%, 99.82\%,  95.68\%,  95.64\%,  91.65\%, respectively, for  OD segmentation, and 91.44\%, 99.87\%,  99.83\%, 85.67\%, 87.48\%, 78.09\%, respectively, for OC segmentation. Further, we perform two-stage glaucoma severity grading using the cup-to-disc ratio (CDR) computed based on the obtained OD/OC segmentation. The superior segmentation performance of RED-RCNN over MRCNN translates to higher accuracy in glaucoma severity grading.
\end{abstract}


%
\IEEEpeerreviewmaketitle

\section{Introduction}
\thispagestyle{empty}
%
%
%
%
\IEEEPARstart{G}{laucoma} is a leading cause of blindness caused by an irreparable and irreversible damage to the optic nerve head (ONH) due to increased intraocular pressure \cite{garway1998quantitative}. Millions of neuronal cells of the visual network converge at the ONH and transmit the visual information from the photoreceptors to the higher levels of processing in the brain. The optic disc (OD) is a two-dimensional representation of the ONH in a retinal fundus image. In the case of glaucoma, the neuroretinal rim tissue gets progressively damaged, which is characterized by {\it cupping} or a {\it depression} on the OD. The depressed region --- termed as the optic cup (OC) --- appears as a bright region originating from the ONH. Accurate segmentation of OD and OC is crucial for the assessment of glaucoma severity. A fast, automated, and accurate segmentation of the OD and OC would enable medical experts to provide reliable diagnosis. Typically, severity assessment of glaucoma is carried out based on the calculation of the cup-to-disc ratio (CDR) or Inferior/Superior/Nasal/Temporal (ISNT) ratio compared against the disc-damage-likelihood scale (DDLS) \cite{henderer2006disc}.\\ 
\begin{figure}[t]
    \centering
    \includegraphics[width = 0.6\linewidth]{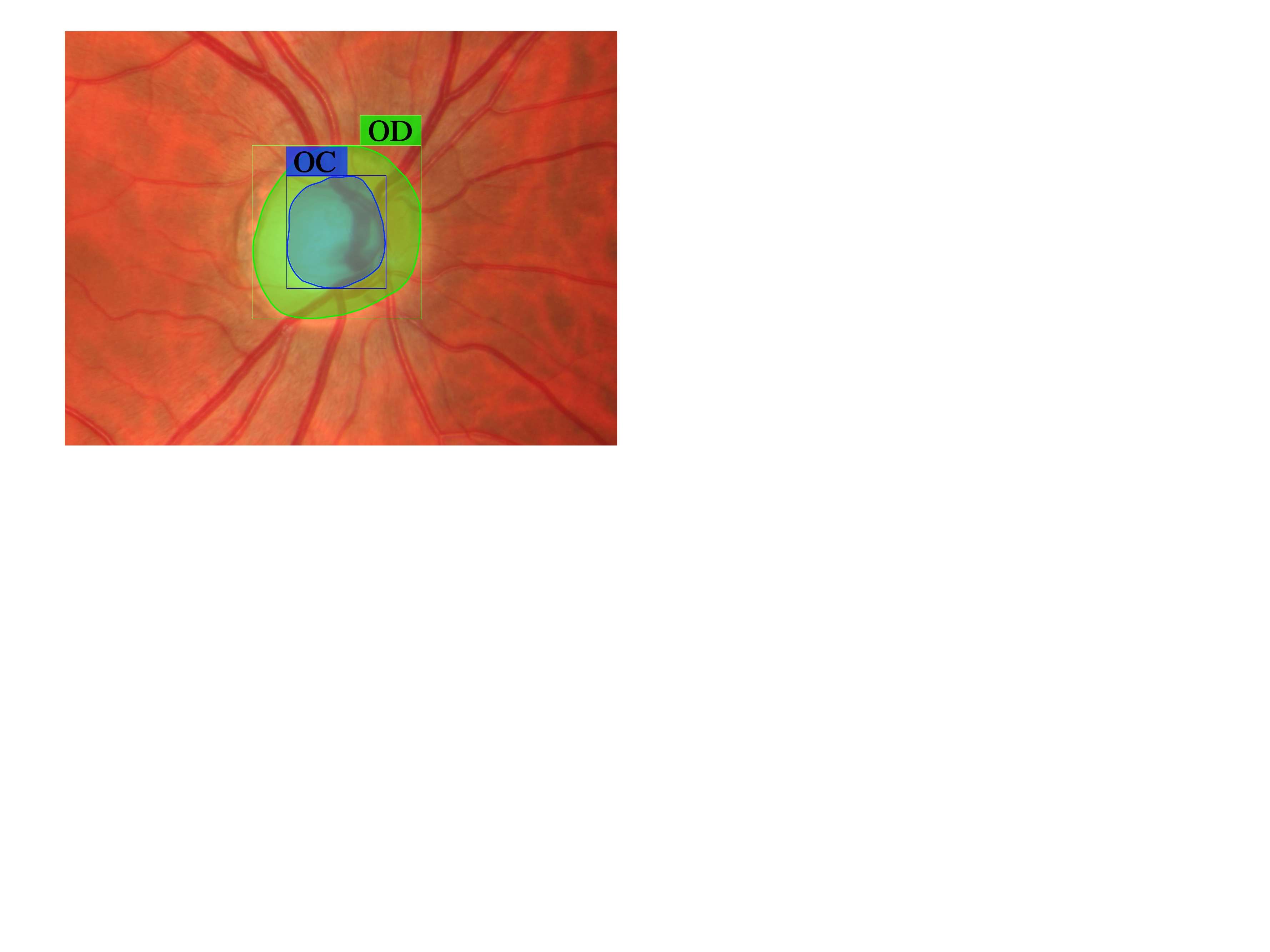}
    \caption{\protect\rgbsymbol\, Simultaneous OD \& OC segmentation using the proposed RED-RCNN.}
    \label{title fig}
\end{figure} 
\indent {\it This paper}: We employ a deep learning approach for fast, simultaneous, and accurate segmentation of OD and OC without imposing shape constraints. Neural networks employed  for segmentation typically require a ground-truth for training, which is often provided by an expert. In cases where multiple expert outlines are available, we need a method for combining them into a single robust outline by taking into account inter-expert variability. We develop a median-based approach for fusing multiple expert annotations into a single one, in order to train the segmentation networks. The starting point for our network architecture is the mask region convolutional neural network (MRCNN), which has proven to be successful for the task of image segmentation on standard datasets \cite{mrcnn}. We show that the MRCNN could be improved by incorporating a residual encoder-decoder network (REDNet) in the mask branch instead of a fully convolutional network (FCN) \cite{rednet}. The new architecture, referred to as RED-RCNN, gives significantly superior results. The details are deferred until Sec.~\ref{sec_proposed_approach}, but an illustration that gives a flavor of the accuracy of segmentation obtained using RED-RCNN is shown in Fig.~\ref{title fig}. The validations are reported on several standard publicly available datasets and  comparisons are provided with state-of-the-art algorithms. The RED-RCNN consistently outperforms the state-of-the-art on almost all the datasets. A two-stage glaucoma analysis on the REFUGE dataset using the CDR parameter shows that RED-RCNN has a superior classification accuracy than MRCNN.

\section{Related Literature} 
\noindent While some works in the literature have focused exclusively on the segmentation of the OD or OC, the recent ones have focused on the simultaneous segmentation of OD and OC. Hence, we classify the related literature under these headings.
\subsection{Optic Disc Segmentation}
\noindent Lowell et al. \cite{lowell2004optic} employed an active contour model to detect the OD based on local image gradients. Abramoff et al. used a pixel feature classification technique to segment the OD into cup, rim, and background from colour stereo photographs \cite{abramoff2007automated}. Aquino et al. employed a template-based approach using feature extraction followed by the circular Hough transform to extract the disc boundary \cite{aquino2010detecting}. Shijian designed a transform that estimates the OD center and its boundary based on image variations \cite{lu2011accurate}, while  Dey et al. used a normalized cross-correlation with a template for disc localization followed by affine snakes in a gradient vector field for OD segmentation \cite{dey2019automatic}. Cardoso et al. presented a model using a combination of the  circular Hough transform and multiscale energy filter for localization and 3-D roughness index value and morphology in conjunction with a fuzzy set representation \cite{dos2019automatic}. The advent of deep learning led to a shift from the use of image processing techniques to learning-based models with a significant improvement in the segmentation performance. Maninis et al. designed a CNN that uses a base-shared network and a specialized layer to segment the OD and demonstrated that the results obtained surpassed human performance \cite{DRIU-2016}. \textcolor{uptxt}{ Mohan et al. presented a Fine-Net CNN model that produces semantic OD boundaries \cite{dm1} and  augmented the results by cascading the Fine-Net CNN with a prior-CNN reported in \cite{dm2}}. Recently, Wang et al. reported improved segmentation performance by augmenting the vanilla U-Net with a fully convolutional network (FCN) \cite{bi2019automated}.

\subsection{Optic Cup Segmentation}
\indent Segmenting the OC is a more challenging task due to the low-contrast boundary. Xu et al. proposed a learning framework operating on a sliding window approach to localize the optic cup \cite{xu2011sliding}. The sliding windows generate candidate samples that are ranked using a support vector regression model employing a radial-basis kernel. The final detection is performed using nonmaximal suppression. Recently, Yang et al. proposed a technique to minimize the influence of blood vessels on the segmentation by using an inpainting model for filling the blood vessels within the OD region \cite{yang2018efficient}. In another attempt, Xu et al. employed an unsupervised approach for the classification of OC {\it superpixels} within a low-rank representation framework \cite{rehman2019multi}. 

\subsection{Optic Disc and Optic Cup Segmentation}
\indent In general, the features that lead to a reliable OD segmentation are different from those that give an accurate OC segmentation. Hence, several works have used differential features for each segmentation task. Joshi et al. used a multidimensional image representation obtained from the colour and texture features to provide a robust outline of the OD \cite{joshi2011optic}. Subsequently, the OC is segmented based on vessel bends and pallor information. Cheng et al. proposed a superpixel classification approach for OD and OC segmentation that accounted for the quality of segmentation through a self-assessment reliability score \cite{cheng2013superpixel}. Zheng et al. integrated OD and OC segmentation within a graph-cut framework that incorporates priors on the location, shape, and size of OD and OC boundaries for training a  Gaussian mixture model, which is then used to estimate the posterior probabilities \cite{zheng2013optic}. Sevastopolsky introduced a modified U-Net to sequentially segment the OD and OC regions  \cite{sevastopolsky2017optic}. Zilly et al. proposed an ensemble learning method to extract OC and OD using CNNs, where an entropy-sampling technique was used to select informative points, and a graph-cut algorithm was employed to obtain the final segmentation result \cite{zilly2017glaucoma}. Singh et al. used the U-Net followed by a multi-scale matching network for segmenting the OD and OC regions \cite{singh2018refuge}. Ghassabi et al.  considered colour, intensity, and orientation contrast computed using Gabor filters as the input to a {\it Winner-take-all} neural network for OD segmentation, and the colour features were provided as input to a self-organizing map neural network for OC segmentation \cite{ghassabi2018unified}. Recently, Edupuganti et al. demonstrated that an FCN based OD-OC segmentation is competitive with the state of the art \cite{edupuganti2018automatic}. Agrawal et al. employed an ensemble of CNNs where, in addition to the visual features, spatial coordinates are fed to the network to improve the segmentation performance \cite{agrawal2018enhanced}. Fu et al. proposed a U-Net backbone with multi-level inputs and polar transformation to improve the segmentation performance \cite{fu2018joint}. \textcolor{uptxt}{Al-Bander et al. proposed FC-DenseNet and extensively validated their OD and OC segmentation methodology \cite{al2018dense} on multiple datsets}. Along similar lines as in  \cite{fu2018joint}, Liu et al. proposed a cartesian-polar network with parallel branches of feature encoders that learn translation-equivariant representations on a cartesian grid and rotation-equivariant representations on a polar grid \cite{liu2019ddnet}. Jiang et al. proposed a two-level strategy using a deep network: a coarse segmentation that fits bounding boxes to the OD and OC, followed by a fine segmentation within the bounding box achieved using a {\it disc-attention module} \cite{jiang2019jointrcnn}.  Yu et al. combined a pretrained ResNet and a U-Net to solve the segmentation problem \cite{yu2019robust}. The use of a pretrained network accelerates learning and achieves high performance. \textcolor{uptxt}{Yin et al. proposed the PM-Net for optic nerve-head localization and used the {\it Pyramid RoIAlign} together with a multi-label head strategy for incorporating priors for improving the OD and OC segmentation performance \cite{yin2019pm}.} Kumar et al. used active discs for segmenting the OD and OC by maximizing the local contrast function \cite{RDR}.

\begin{figure}[!htb]
    \centering
    \includegraphics[width=0.45\textwidth]{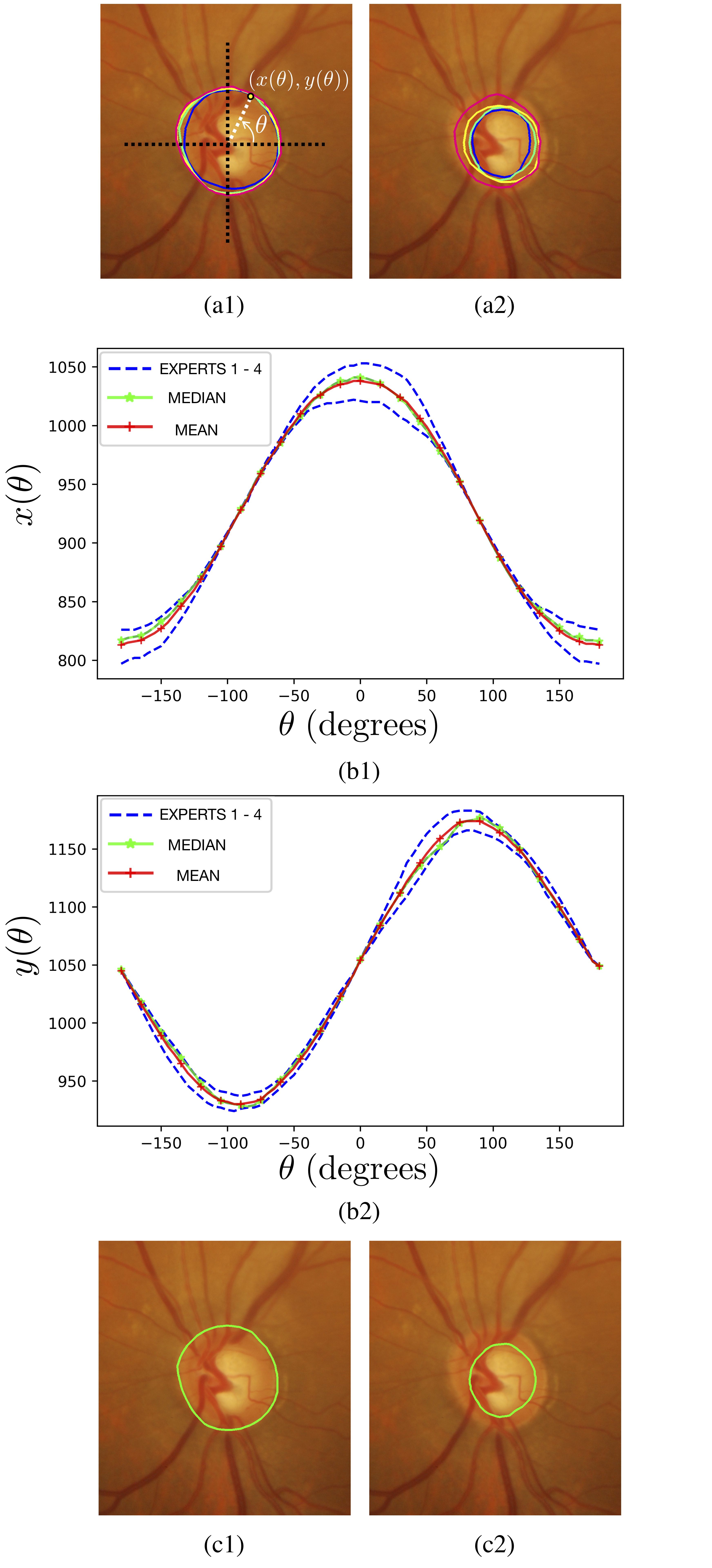}\\
 \caption{\protect\rgbsymbol\, (a1) \& (a2) show the multiple expert annotations for OD and OC, respectively, on a sample image taken from the Drishti-GS dataset \cite{almazroa2018retinal}; (b1) \& (b2) show the $x$ and $y$ coordinate functions, respectively, vs. the polar angle $\theta$ (cf. (a1)); (c1) \& (c2) show the median contour (green) for the OD and OC, respectively.} 
 \label{cnt_fuse}
\end{figure}

\begin{figure}[!htb]
    \centering
    \includegraphics[width=0.45\textwidth]{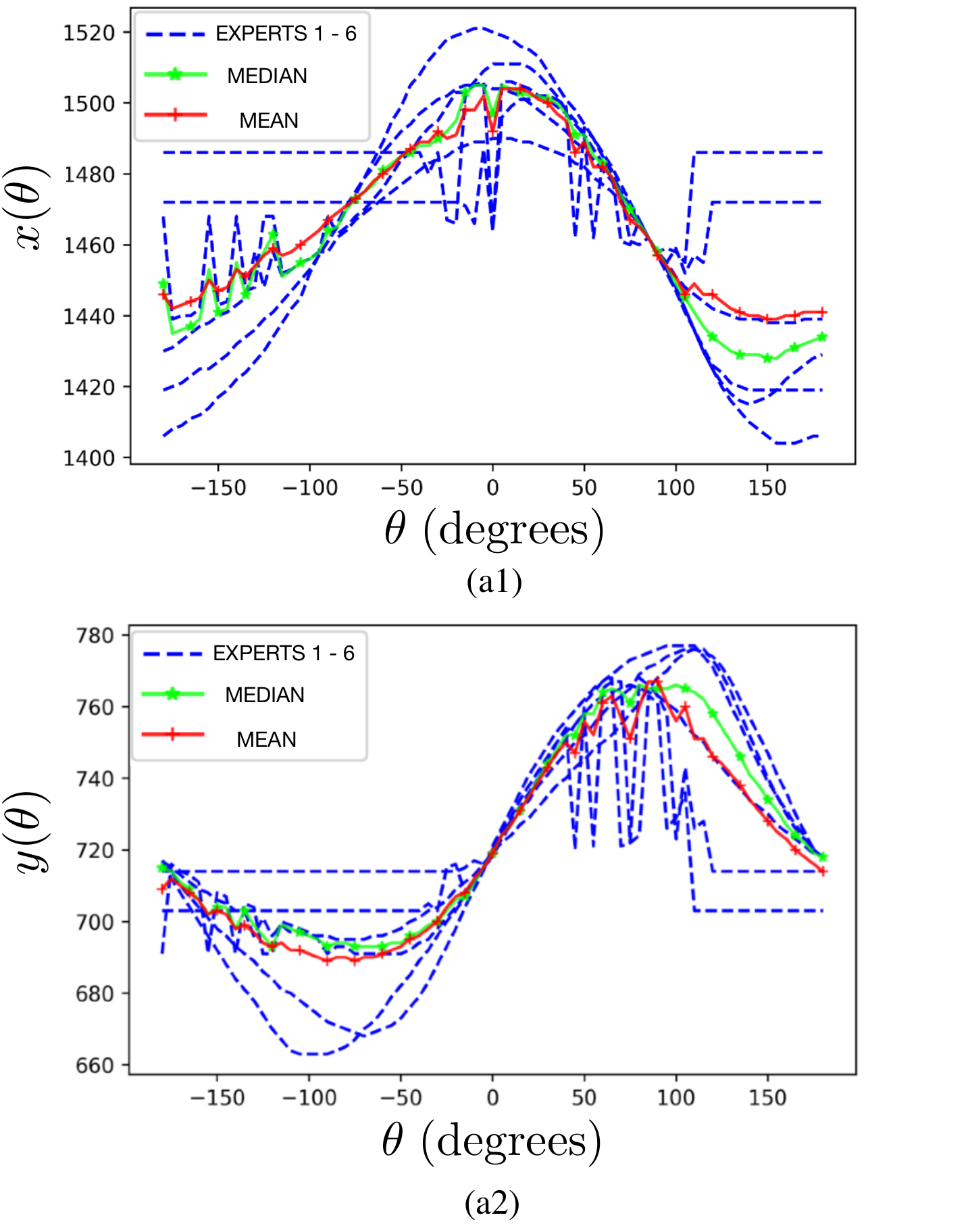}
      \caption{\protect\rgbsymbol\, The (a) $x$ and (b) $y$ coordinate functions vs. the polar angle $\theta$ obtained from six expert outlines of the OC on a sample image taken from RIGA dataset \cite{almazroa2018retinal}. The median results in a more robust fusion than the mean.}
    \label{mnmd}
\end{figure}

\section{Robust Fusion of Multiple Expert Annotations}
\indent Training a neural network to perform OD and OC segmentation typically requires ground-truth annotations from expert ophthalmologists and in many cases, more than one such annotation may be available. Of the publicly available datasets, Drishti-GS \cite{sivaswamy2015comprehensive}, RIGA \cite{almazroa2018retinal}, and REFUGE \cite{refuge@2018} contain both disc and cup outlines, out of which Drishti-GS and RIGA contain annotations from multiple clinical experts. Neural networks typically require a single ground-truth for training, which calls for a robust way of fusing the multiple annotations into a single one. Taking the intersection of the multiple annotations may be conservative, whereas taking the union would bias the ground-truth toward annotations covering a larger area, which is not preferred either. We develop a method for fusing annotations in a way that is robust to inter-expert variability.\\
\indent Figs.~\ref{cnt_fuse}(a1) \& (a2) show multiple expert outlines of the optic disc and cup, respectively. Figs.~\ref{cnt_fuse}(b1) \& (b2) show plots of the $x$ and $y$ coordinates of the disc outlines as a function of the polar angle $\theta$ (Fig.~\ref{cnt_fuse}(a1)). A similar plot for the cup outlines is shown in Figs. ~\ref{mnmd}(a) \& (b). 
We observe that the median is more robust to inter-expert variability and is a better candidate than the mean for fusing multiple annotations. Figs.~\ref{cnt_fuse}(c1) \& (c2) show the median ground-truth outlines of the OD and OC, respectively.

\section{The Proposed Approach} \label{sec_proposed_approach}
\indent The proposed residual encoder-decoder region CNN (RED-RCNN) approach is based on the mask region convolutional neural network (MRCNN) and the residual encoder-decoder network (REDNet) \cite{rednet}, neither of which were used for OD/OC segmentation. We highlight the features of the MRCNN and REDNet and then proceed with the proposed approach.

\subsection{Mask Region Convolutional Neural Network (MRCNN)}
\begin{figure*}[t!]
      \centering
    \includegraphics[width=0.875\textwidth]{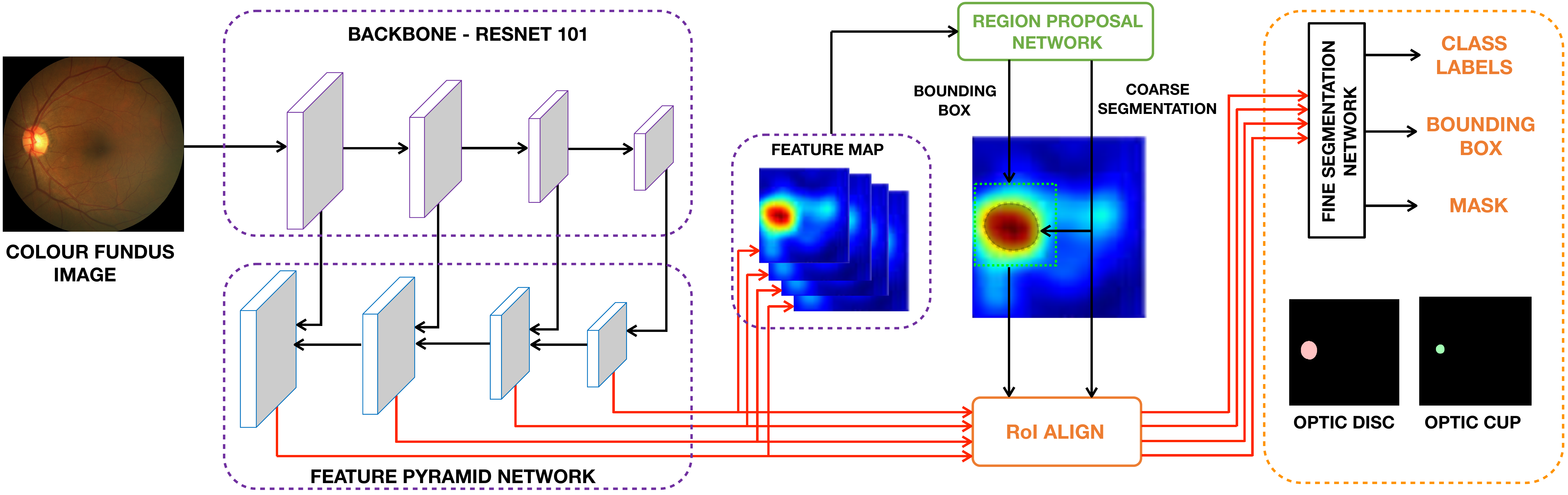}
     \caption{\protect\rgbsymbol\, Workflow in an MRCNN detailing the simultaneous OD and OC segmentation.} \label{mcnbk}
\end{figure*}

\begin{figure*}[h]
    \centering
    $\begin{array}{cc}
    \hspace{1mm}\includegraphics[width=0.45\textwidth]{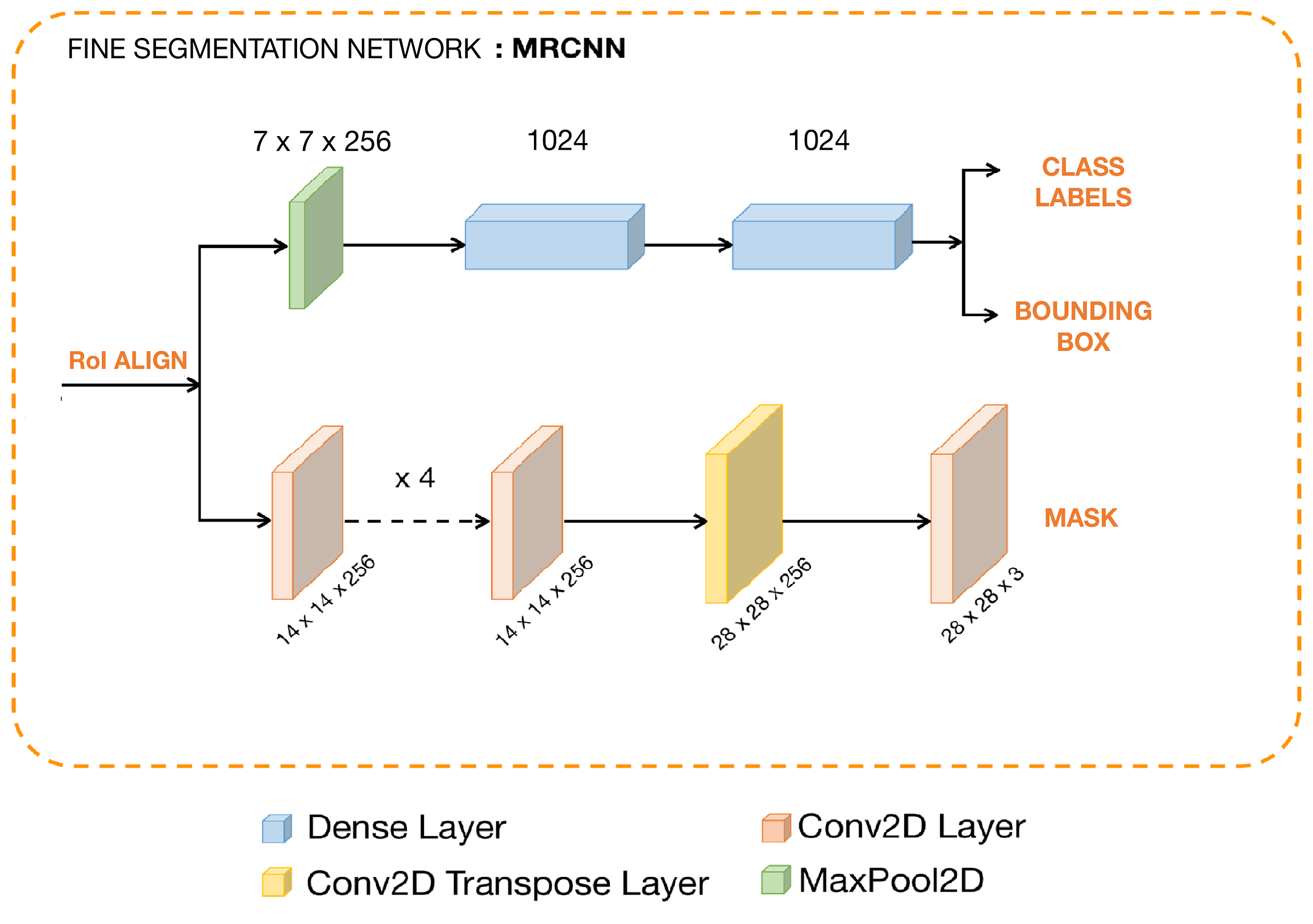}& \hspace{1mm}\includegraphics[width=0.45\textwidth]{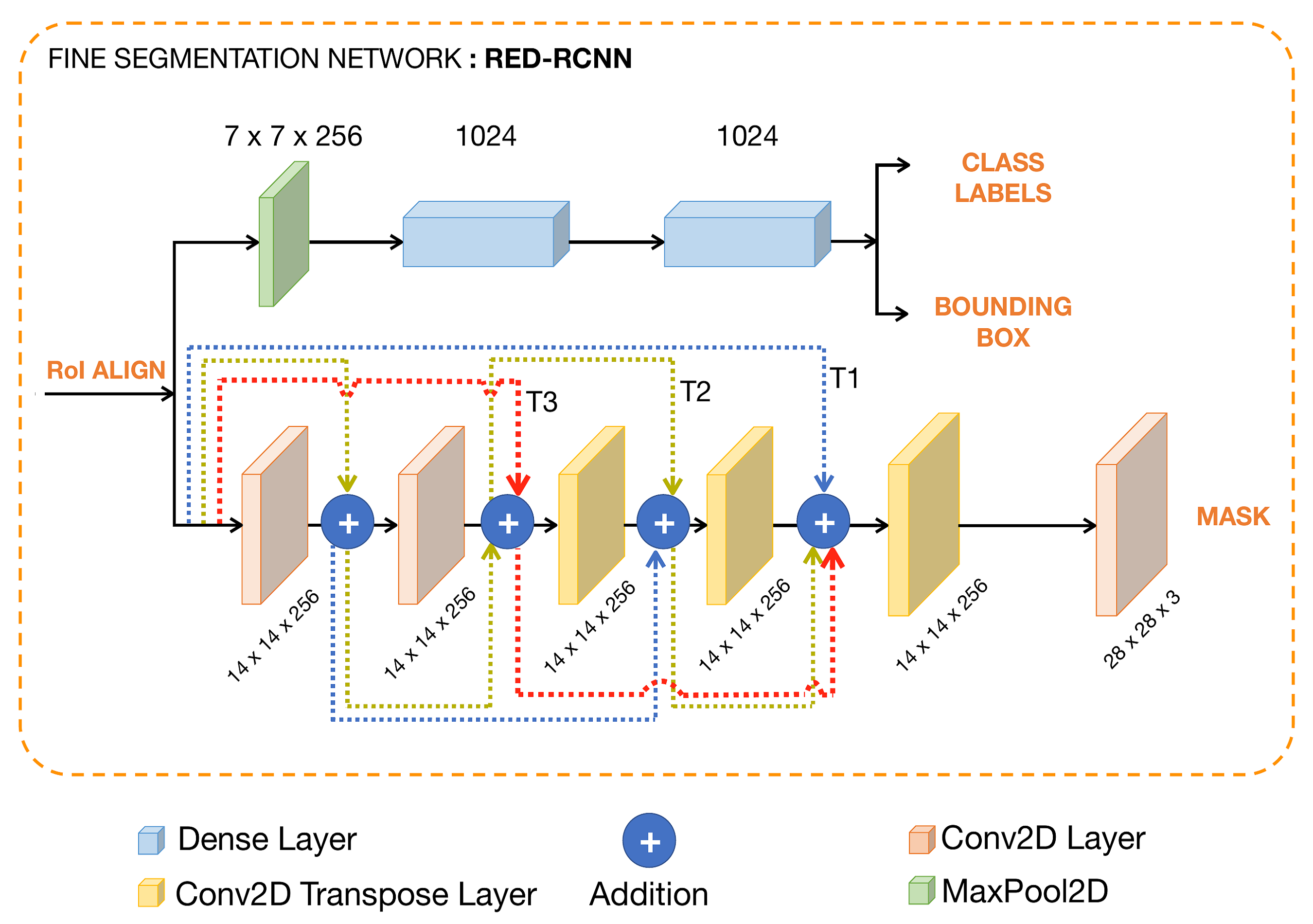}\\
    \mbox{(a)}&\mbox{(b)}\\
    \end{array}$
    \caption{\protect\rgbsymbol\, The Fine Segmentation Network of (a) the MRCNN, and (b) \textcolor{uptxt}{the proposed RED-RCNN, with Type-1 (T1), Type-2 (T2), Type-3 (T3) being different types of skip connections.}}
    \label{fineseg}
\end{figure*} 

\indent Mask R-CNN (MRCNN) \cite{mrcnn} is a deep neural network used for instance segmentation of objects, i.e., it can differentiate objects in an image through detection and delineation of the regions of interest. MRCNN is an extension of the Faster R-CNN (FRCNN) \cite{frcnn}. A block diagram of MRCNN is shown in Fig.~\ref{mcnbk}. It has a multistage architecture comprised of three main blocks: a backbone network in tandem with a feature pyramid network (FPN) that together perform multiscale feature extraction on a colour fundus image; a region proposal network (RPN) that performs coarse segmentation on the multiscale feature maps; and a fine segmentation network that refines the segmentation map obtained from the preceding two blocks.\\
\indent The backbone consists of top-down, bottom-up and lateral connections. The top-down connection generates feature maps. The bottom-up connection is usually a CNN that extracts features from the input image. Lateral connections allow for additions and convolutions between the top-down and bottom-up pathways. The RPN scans the feature maps and determines the regions that have a high probability of containing an object. It outputs a region of interest (RoI) containing a bounding box and a coarse segmentation map. The RoI is aligned with the multiscale feature maps and fed to a fine segmentation network (FSN), which in turn outputs the class labels, a refined bounding box, and a fine segmentation mask. 

\subsection{Residual Encoder-Decoder Network (REDNet)}
\indent The accuracy of segmentation can be improved if the context of the pixel with its neighbors is also taken into account. There are several segmentation approaches that are based on fully convolutional networks (FCNs) that downsample the feature maps using a max-pool layer after each convolution. Badrinarayanan et al. \cite{segnet} pointed out that max-pooling results in the loss of location information, which is crucial for segmentation and instead suggested using pooling-indices to keep track of the locations of the max-values at the expense of more memory for storing the locations. As an alternative to storing the indices, Noh et al. \cite{dconvnet} used a deconvolution layer that learns to optimally upsample the input. Mao et al. \cite{rednet} designed a residual encoder-decoder network by eliminating the max-pool layer and used deconvolutions along with residual skip connections to get more accurate segmentation maps.

\subsection{Residual Encoder-Decoder Region CNN (RED-RCNN)}
\label{skip}
\indent Although MRCNN has been used for several object segmentation tasks, it has never been applied to fundus images. We begin with employing the MRCNN for simultaneous OD and OC segmentation, the results of which will be reported in the subsequent section. The results show that MRCNN by itself is quite competitive with the state-of-the-art approaches. Further, we leverage the superior segmentation performance of the REDNet and employ it in the mask branch of the fine segmentation network. Also, the standard MRCNN uses a single deconvolutional layer in the final stage of segmentation, essentially for upsampling the segmentation map to match the size of the input image. On the other hand, REDNet has a deconvolutional layer in the decoder for every convolutional layer in the encoder, thus giving rise to a learnt stage-wise upsampling mechanism. Drozdzal et al. \cite{imp_skip} highlighted the importance of skip connections in deep networks, showing better gradient flow, resulting in faster convergence and superior performance. We experimented with three types of skip connections: \textcolor{uptxt}{ Type-1 (T1)}: between two consecutive convolutional layers; \textcolor{uptxt}{Type-2 (T2)}: between every pair of convolutional and deconvolutional layers; and \textcolor{uptxt}{Type-3 (T3)}: between a convolutional block and a deconvolutional block (each block comprising two layers). We found that T3 (cf. Fig.~\ref{fineseg}(b)) gave the best performance, which is further explained in Sec.~\ref{skip_con}. Therefore, we used REDNet with Type-3 skip connections in the mask branch of the MRCNN. Henceforth, we shall refer to MRCNN with REDNet as RED-RCNN. Effectively, the RED-RCNN comprises the MRCNN in Fig.~\ref{mcnbk} but with the fine segmentation network as given in Fig.~\ref{fineseg}(b). A comprehensive evaluation and comparison of the segmentation performance of MRCNN and RED-RCNN vis-\`a-vis the state of the art will be reported in Sec.~\ref{exptval}.

\begin{table}[t!]
\caption{An overview of the datasets used for validation.}
\centering
\resizebox{\columnwidth}{!}{
\begin{tabular}{@{} c | c | c | c | c | c @{} }
\hline \hline
\textbf{Dataset} & \textbf{\# Images} & \textbf{Resolution} & \textbf{OD} & \textbf{OC} & \textbf{\# Experts} \\
& \textbf{used}& &\textbf{ground-truth} &\textbf{ground-truth} & \\
\hline
Drishti-GS \cite{sivaswamy2015comprehensive}& 101 & 2896 $\times$ 1944 & \checkmark & \checkmark & 4 \\
\hline
REFUGE \cite{refuge@2018}& \textcolor{uptxt}{800} & 2124 $\times$ 2056  & \shortstack{\checkmark} & \checkmark & 1 \\
&&1634 $\times$ 1634 & & & \\
\hline
RIGA \cite{almazroa2018retinal}& 750 & varying & \checkmark & \checkmark & 6 \\
\hline
IDRiD \cite{idrid}& 81 & 4288 $\times$ 2848 & \checkmark & $\times$ & 1 \\
\hline
DRIONS-DB \cite{carmona2008identification}& \textcolor{uptxt}{50} & 600 $\times$ 400 & \checkmark & $\times$ & 1 \\
\hline
MESSIDOR \cite{decenciere_feedback_2014}& 1200 & 2240 $\times$ 1488 & \checkmark & $\times$ & 1 \\
\hline
\hline
\end{tabular}
}
\label{datasets}
\end{table}
\subsection{Loss Function for the Mask Branch}
\indent The original proponents of MRCNN used binary cross-entropy ($BCE$) as the loss function for the mask branch. Since the segmentation performance is typically quantified using the intersection over union ($IoU$), we prefer to augment the $BCE$ loss with the $IoU$ for training the network as it is likely to improve the segmentation performance. However, since $IoU$ is not differentiable, we employ a differentiable approximation to the $IoU$ proposed by Rahman and Wang \cite{IOU}, given by
\begin{equation}{\widehat {IoU}}=\frac{\sum_{ij}X_{ij}Y_{ij}}{\sum_{ij} X_{ij}+Y_{ij}-X_{ij}Y_{ij}},\end{equation}
where $ij$ denotes the pixel index, $X_{ij}$ is the prediction, and $Y_{ij}$ is the ground-truth.
We consider a convex combination of $BCE$ and ${\widehat {IoU}}$: 
\begin{equation}
\mathcal{L}_{mask} = \alpha\,BCE + (1-\alpha)\,{\widehat {IoU}}, \alpha \in [0, 1],
\label{augloss}
\end{equation}
for training the mask branch in both MRCNN and RED-RCNN. The choice of $\alpha$ will be explained in the next section. The loss functions for obtaining the class labels (sparse softmax cross-entropy) and the bounding box (smooth L1 loss) are retained as proposed in the original MRCNN paper.

\begin{figure*}[t]
  \centering
  \hspace{-0.4cm}
  \begin{minipage}{0.99\textwidth}
  \centering

 $\begin{array}{cccccc}
    \includegraphics[width=0.149\textwidth]{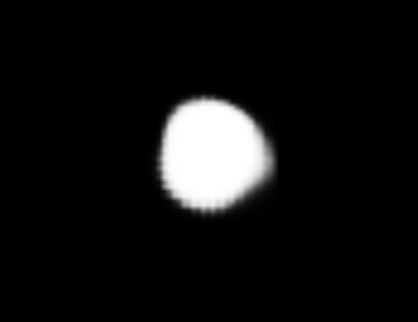} &
    \includegraphics[width=0.149\textwidth]{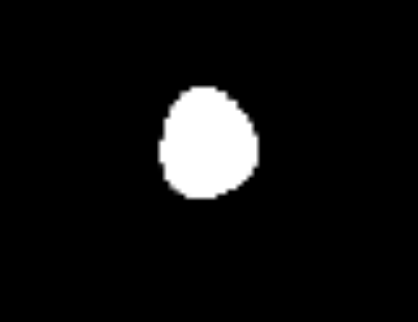} &
    \includegraphics[width=0.149\textwidth]{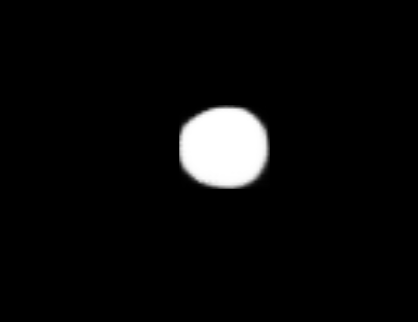} &
    \includegraphics[width=0.149\textwidth]{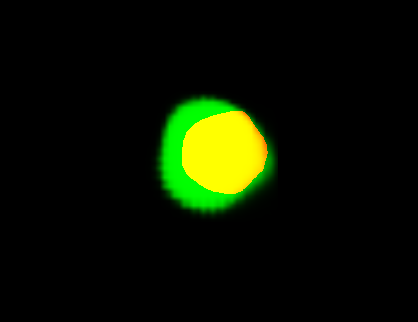} &
    \includegraphics[width=0.149\textwidth]{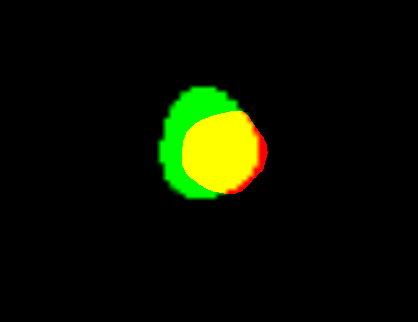} &
    \includegraphics[width=0.149\textwidth]{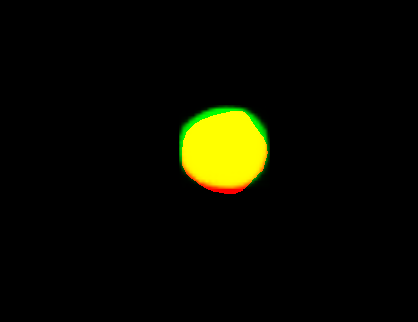}\\
    \mbox{(a1) $BCE$}&\mbox{(a2) ${\widehat{IoU}}$}&\mbox{(a3) Aug. Loss} & \mbox{(b1)}&\mbox{(b2)}&\mbox{(b3)}\\
    \end{array}$
    \label{loss1}
    \end{minipage}
    \vspace{0.5cm}
    \begin{minipage}{0.99\textwidth}
        \centering
        \vspace{0.3cm}
        $\begin{array}{ccc}
        \hspace{-0.15cm}
        \includegraphics[width=0.339\textwidth]{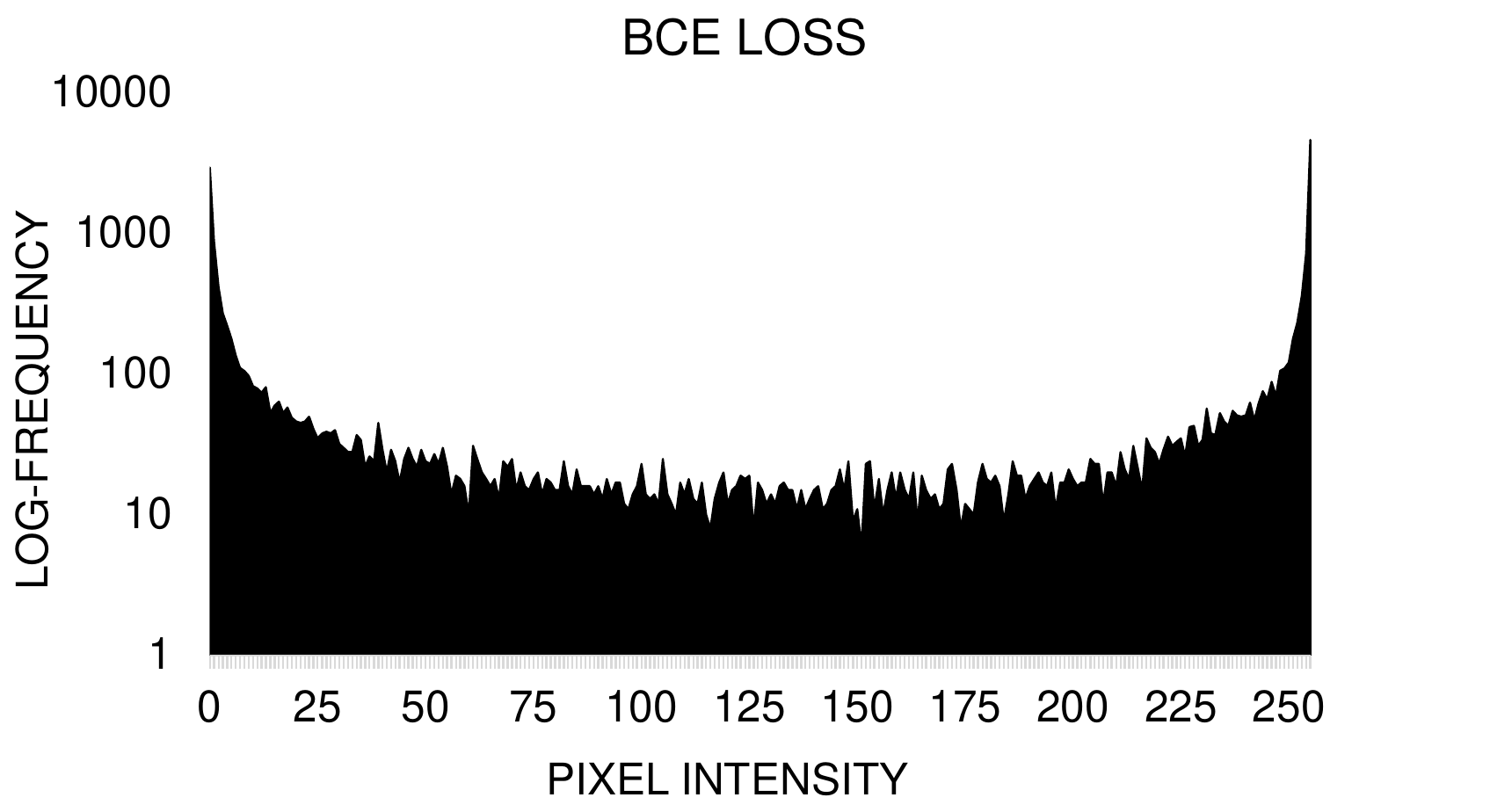} &
        \hspace{-0.36cm}
        \includegraphics[width=0.339\textwidth]{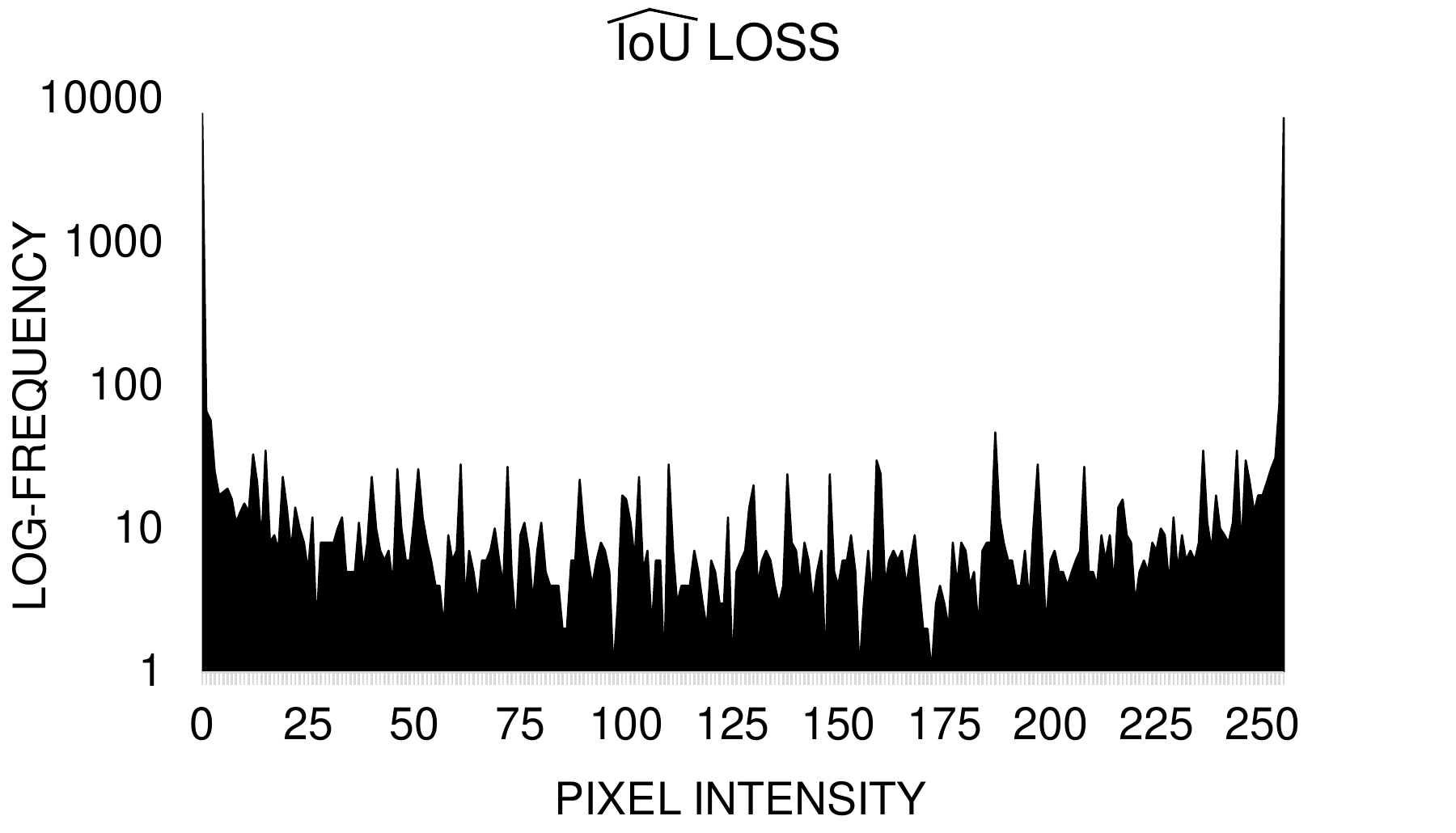} &
        \hspace{-0.29cm}
        \includegraphics[width=0.339\textwidth]{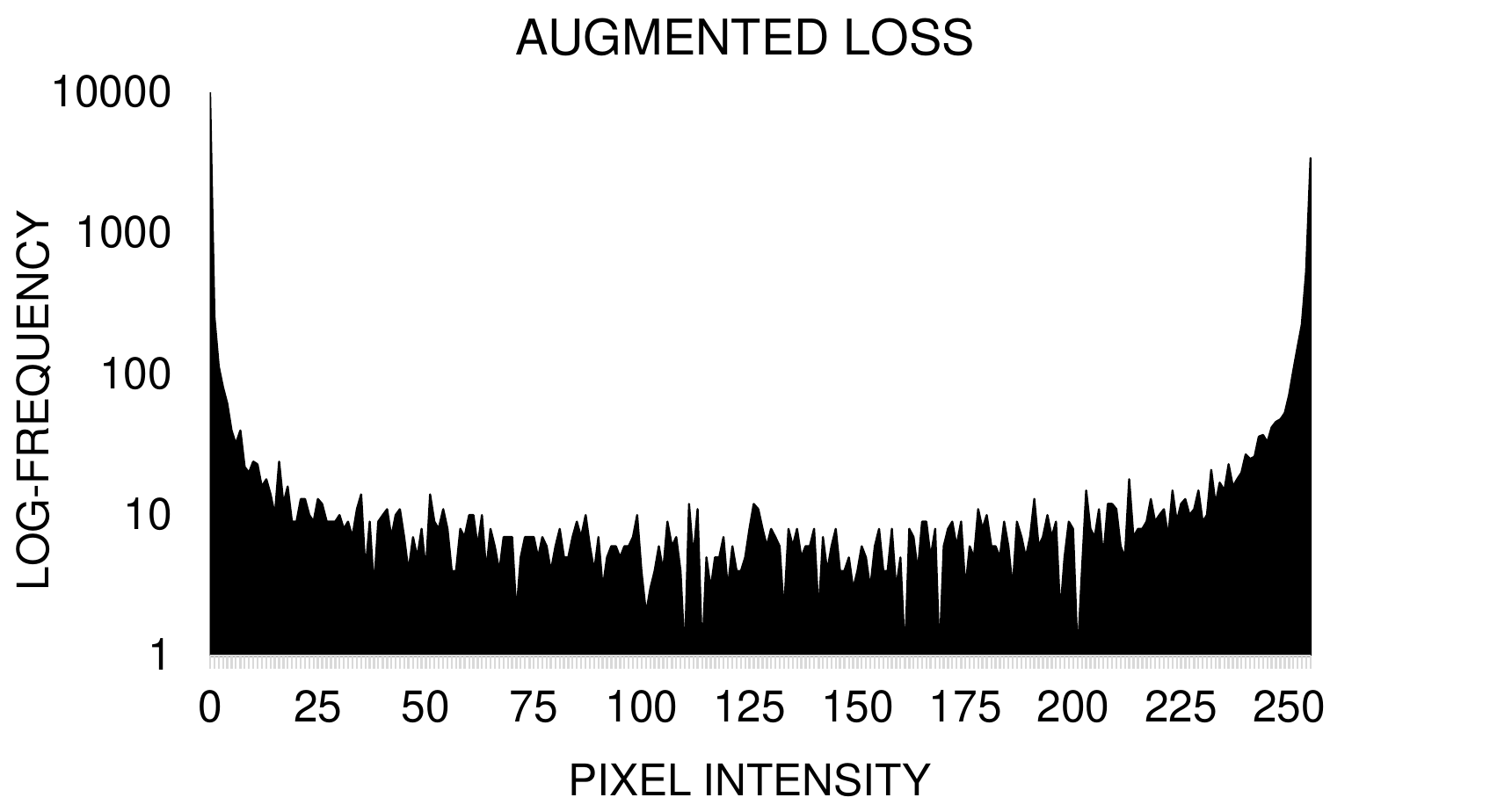} \\
        \mbox{(d)} & \mbox{(e)} & \mbox{(f)}\\
        \label{Loss2}
        \end{array}$
    \end{minipage}
    \vspace{-0.7cm}
\caption{\protect\rgbsymbol\, \textcolor{uptxt}{To illustrate the advantage of the augmented loss function: (a1), (a2), and (a3) show the OC predictions obtained using \textit{BCE}, \textit{$\widehat{IoU}$}, and the augmented loss, respectively; (b1), (b2), and (b3) characterize the accuracy of the predictions shown in (a1), (a2), and (a3), respectively, in comparison with the ground-truth. The color coding is as follows: yellow: true positive; red: false negative; green: false positive. (b3) shows the highest true positives and the lowest false positives and false negatives. The subplots (d), (e), and (f) show the histograms (y-axis on log-scale) of pixel intensities in the predictions shown in (a1), (a2), and (a3), respectively; $0$ and $255$ correspond to the background and foreground, respectively.}}    
\label{Loss_FN}
\end{figure*}


\section{Experimental Validation}
\label{exptval}
\subsection{Datasets}
\noindent We carried out extensive validation on publicly available datasets: Drishti-GS\cite{sivaswamy2015comprehensive}, REFUGE \cite{refuge@2018}, RIGA \cite{almazroa2018retinal}, IDRiD \cite{idrid}, DRIONS-DB \cite{carmona2008identification}, and MESSIDOR \cite{decenciere_feedback_2014}. A summary of the number of images used, image resolution, availability of the OD and OC ground-truth, and the number of expert annotations, for each of the datasets is shown in Table~\ref{datasets}.

\subsection{Network Training}
\label{net_tr}
\indent The implementation and hyper-parameter tuning of RED-RCNN is based on that of MRCNN \cite{matterport_maskrcnn}. As in the case of MRCNN, an RoI is considered positive only if its IoU is greater than 0.5. We use ResNet101-FPN \cite{resnet101} as the backbone network. All the images are resized such that the maximum row/column dimensions do not exceed 1024 while maintaining the image aspect-ratio by zero-padding appropriately. For the purpose of initialization, we use MRCNN pretrained on the COCO dataset, which is also available at \cite{matterport_maskrcnn}. The training is performed on two Nvidia RTX 2080Ti GPUs using ADAM optimizer \cite{kingma2014adam} with a learning rate of 0.01, momentum of 0.9, and a weight decay of 0.0001, for 450 iterations. \\
\indent A fundus image has only one instance of the OD and OC. Hence, the maximum number of ground-truth instances in the MRCNN is set to 1 and the number of proposals generated by RPN is reduced to 32. While testing, we set 0.9 as the detection threshold in the fine segmentation network.\\
\indent We used Drishti-GS, REFUGE, and RIGA datasets for training as they contain both OD and OC outlines. During the training phase, we choose a random $70\%-20\%$ split of the data, where the $70\%$ part is used for training and the $20\%$ part is used for cross-validation. The remaining $10\%$ is not seen during training. Each iteration is composed of 100 passes of random image batches of size eight from the training set. After every iteration, the model is cross-validated on the $20\%$ data and the model performance (the loss function values) and the corresponding weights are recorded. Both MRCNN and RED-RCNN models are trained for 450 iterations. Once the training phase is complete, we selected the weights that gave the best performance on the cross-validation set and then validated the corresponding model on the $10\%$ unseen data.

\begin{figure*}[t]
    \centering
    $\begin{array}{cc}
    \hspace{-0.3cm}
     \includegraphics[width=0.48\textwidth]{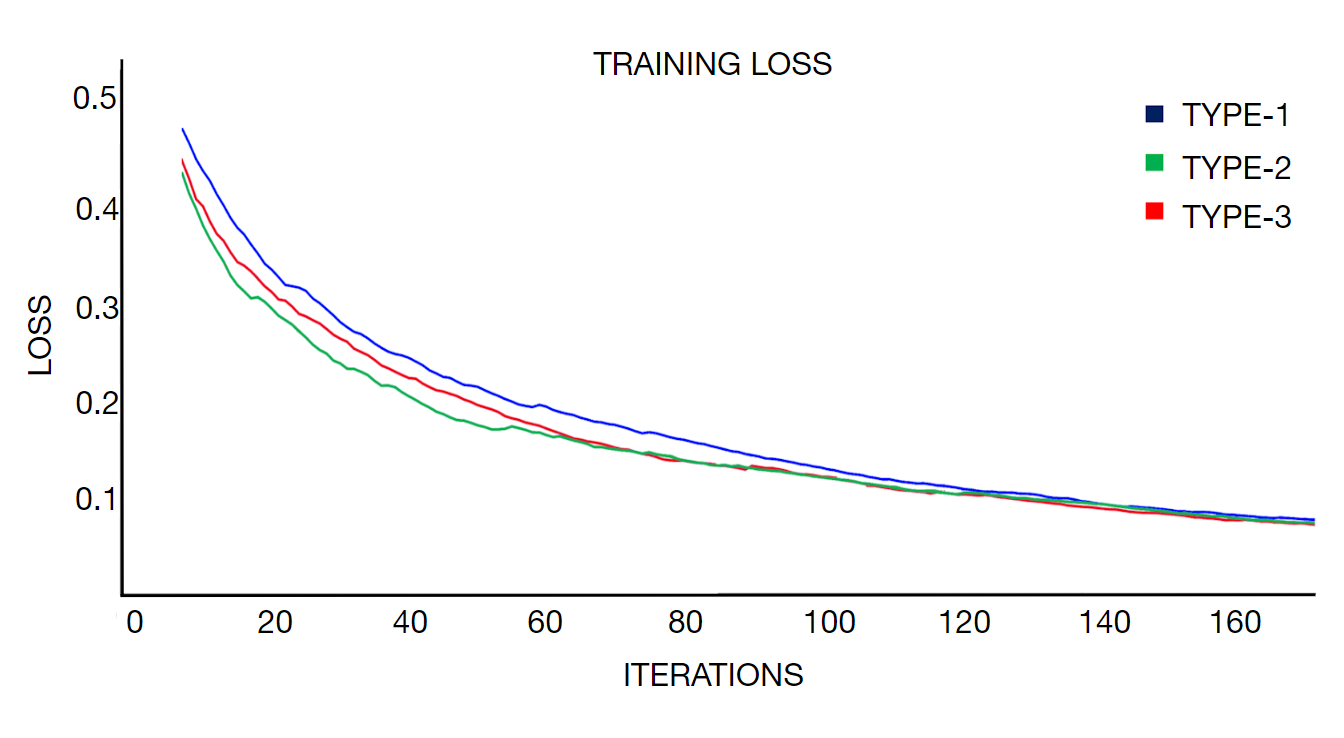}&
     \hspace{-0.33cm} \includegraphics[width=0.48\textwidth]{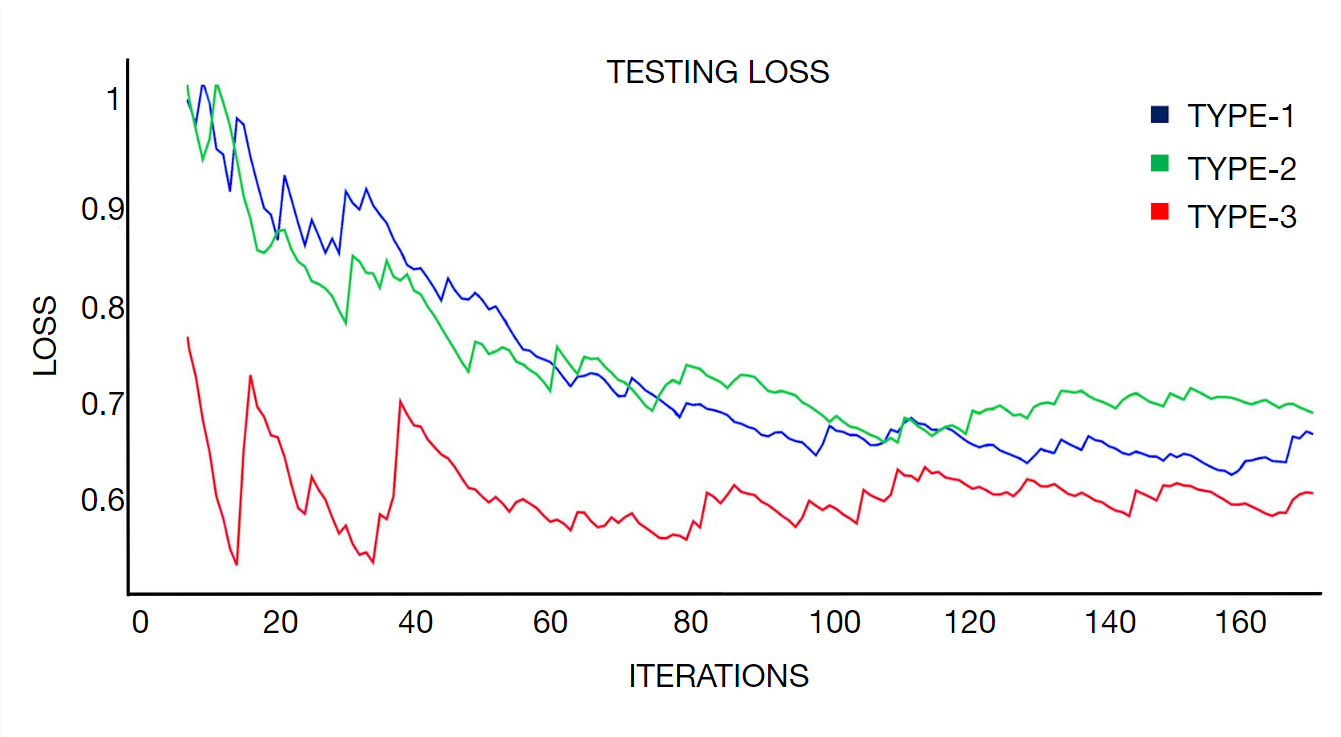}\\
      \mbox{(a)} & \mbox{(b)}\\
    \end{array}$
\caption{\protect\rgbsymbol\, \textcolor{uptxt}{A comparison of the augmented loss on (a) the training data, and (b) the test data, for various types of skip connections. Although all three types result in comparable training losses upon convergence, the testing loss is the least in the case of Type-3 skip connection.}}
\label{Skip_comp}
\end{figure*}

\textcolor{uptxt}{
\subsection{Choice of $\alpha$ and the Significance of the Augmented Loss}
\indent Setting $\alpha$ to either 0 or 1 reduces the loss in Eq.~(\ref{augloss}) to the standard \textit{$\widehat{IoU}$} or \textit{BCE} loss functions, respectively. For most neural networks that perform segmentation, accurate classification of the OC boundary within the OD still remains a challenge. In the final stages of training a network, we observed that the predictions along the boundaries of the RoI have varying probabilities, whereas those within the RoI have more homogeneous probabilities. Using the \textit{BCE} loss alone results in negligible gradients for learning in the final stages, which is because large portions of the image except at the boundaries have nearly homogeneous predictions. On the other hand, using \textit{$\widehat{IoU}$} loss would be beneficial but it requires a large number of iterations to achieve convergence. Hence, we considered a convex combination of the two losses. To begin with, the parameter $\alpha$ was initialised at a high value and then decreased progressively. Consequently, the weight $(1 - \alpha)$ assigned to \textit{$\widehat{IoU}$} went up in the later stages, which helped in generating sufficient gradients to facilitate learning and to classify the boundary pixels with a higher accuracy. In our implementation, the parameter $\alpha$ was initialized to 0.7, and decremented by 0.01 once in every ten iterations. The initial value $\alpha=0.7$ was fixed experimentally.\\
\indent Using the augmented loss also serves an important purpose. The network outputs after scaling have pixel intensities distributed between 0 and 255. Segmentation of the RoI is performed by setting a single threshold on the predicted outputs, for classifying the pixels as foreground or background. A single threshold was found to be suboptimal due to the misclassification of the boundary pixels along the RoI. Augmenting the \textit{BCE} loss with \textit{$\widehat{IoU}$} reduced the rate of misclassificaton of the boundary pixels.  Figure~\ref{Loss_FN} shows the OC predictions and the true positives, false positives, and false negatives in comparison with the ground-truth. The best results are obtained in the case of the augmented loss function. Figs.~\ref{Loss_FN}(d)-(f) show that the augmentation results in a more decisive action --- the histograms are edge-peaked and fewer pixels have intermediate intensities, thereby reducing the sensitivity to the choice of the single threshold value of $127$.}
\textcolor{uptxt}{
\subsection{Skip Connection}
\label{skip_con} As mentioned in Sec.~\ref{skip}, we consider three types of skip connections in the Residual Encoder-Decoder network. Figure~\ref{Skip_comp}(a) and (b) show the loss values as a function of the iterations on the train and test data, respectively, for the three types of skip connections considered. The figure shows that Type-3 skip connection, which links a convolutional block and the deconvolutional block, gives the best results. Therefore, for all further evaluations of the RED-RCNN, we consider the mask branch with Type-3 skip connection only.}

\begin{table*}[ht]
\centering
\footnotesize
\caption{Training and test data performance of MRCNN and RED-RCNN for OD segmentation. The average values are highlighted on grey background. \textcolor{uptxt}{$^+$Train dataset (1460 images): 91 (Drishti) + 719 (REFUGE) + 650 (RIGA); $^*$Test dataset (190 images): 10 (Drishti) + 81 (REFUGE) + 99 (RIGA).}}
\vspace{0.3cm}
\begin{tabular}{@{}|c|c|c|c|c|c|c|c|@{}}
\hline
\hline
\textbf{Network} & \textbf{Datasets} &\textbf{Sensitivity} & \textbf{Specificity} & \textbf{Accuracy} & \textbf{Precision} & \textbf{Dice} & \textbf{Jaccard} \\ 
&(Total no. of images)&&&&&&\\
\hline
& ($+$) Drishti-GS  & 0.9471&	0.9995&	0.9979&	0.9845	&0.9654&	0.9331  \\ 
\textbf{MRCNN}&($+$) RIGA        &0.9431 &	0.9998&	0.999 &	0.9865 &	0.9643 &	0.9310  \\ 
&($+$) REFUGE     &0.9118 &	0.9998 &	0.9983 &	0.9845 &	0.9468 &	0.8989  \\ 
&(1460) & \cellcolor{Gr}0.934	&\cellcolor{Gr}0.9997	&\cellcolor{Gr}0.9984	&\cellcolor{Gr}0.9851&\cellcolor{Gr}	0.9588&\cellcolor{Gr}	0.921\\
\hline
& ($+$) Drishti-GS  & 0.9852 &	0.9982 &	0.9978 &	0.9462	&0.9653 &	0.9329 \\ 
\textbf{RED-RCNN}&($+$) RIGA        &0.9623 &	0.9994 &	0.9989 &	0.9615 &	0.9619 &	0.9266 \\ 
&($+$) REFUGE     &0.9594 &	0.9996 &	0.999 &	0.9779 &	0.9686 &	0.9392  \\ 
&(1460) &\cellcolor{Gr} 0.9689 &\cellcolor{Gr}	0.999 &\cellcolor{Gr}	0.9985 &\cellcolor{Gr}	0.9618 &\cellcolor{Gr}	0.9652 &\cellcolor{Gr}	0.9329 \\
\hline
& ($*$) Drishti-GS  & 0.8362 &	0.9997 &	0.9949 &	0.99 &	0.9066 &	0.8292 \\ 
\textbf{MRCNN}&($*$) RIGA        &0.9390 &	0.9997 &	0.9988 &	0.9755 &	0.9569 &	0.9174\\ 
&($*$) REFUGE     &0.9192 &	0.9998 &	0.9985 &	0.989 &	0.9528 &	0.9099\\ 
 &(190) &\cellcolor{Gr} 0.8981 &\cellcolor{Gr}	0.9997 &\cellcolor{Gr} 	0.9974 &\cellcolor{Gr}	0.9848 &\cellcolor{Gr}	0.9387 &\cellcolor{Gr}	0.8855 \\
\hline
\textbf{RED-RCNN}& ($*$) Drishti-GS  &0.9839 &	0.9975 &	0.9971 &	0.9216 &	0.9517 &	0.9078 \\ 
$+$: Train dataset  &($*$) RIGA        &0.9623 &	0.9994	&0.9989	 &0.9615 &	0.9619 &	0.9266\\ 
$*$: Test dataset &($*$) REFUGE     &0.9572 &	0.9996 &	0.9989 &	0.9743 &	0.9657 &	0.9336 \\ 
  \cellcolor{Gr} \textbf{Average}&(190) & \cellcolor{Gr}0.9678 &\cellcolor{Gr}	0.9988 &\cellcolor{Gr}	0.9983 &\cellcolor{Gr}	0.9524 &\cellcolor{Gr}	0.9597&\cellcolor{Gr}	0.9226 \\
\hline
\hline
\end{tabular}
\label{trte od}
\end{table*}
\vspace{0.2cm}
\begin{table*}[ht!]
\centering
\footnotesize
\caption{Training and test data performance of MRCNN and RED-RCNN for OC segmentation. The average performance indices are also provided. \textcolor{uptxt}{$^+$Train dataset (1460 images): 91 (Drishti) + 719 (REFUGE) + 650 (RIGA); $^*$Test dataset (190 images): 10 (Drishti) + 81 (REFUGE) + 99 (RIGA).}}
\vspace{0.5cm}
\begin{tabular}{@{}|c|c|c|c|c|c|c|c|@{}}
\hline
\hline
\textbf{Network} & \textbf{Datasets} &\textbf{Sensitivity} & \textbf{Specificity} & \textbf{Accuracy} & \textbf{Precision} & \textbf{Dice} & \textbf{Jaccard} \\ 
&(Total no. of images)&&&&&&\\
\hline
& ($+$) Drishti-GS  & 0.6111 &	0.9999&	0.9934&	0.995&	0.7572&	0.6092  \\ 
\textbf{MRCNN}&($+$) RIGA        &0.8788&	0.9998&	0.9994&	0.949&	0.9125&	0.8391  \\ 
&($+$) REFUGE     &0.8472&	0.9997	&0.999&	0.9162	&0.8804	&0.7863  \\ 
&(1460) &\cellcolor{Gr} 0.7790&\cellcolor{Gr}	0.9998&\cellcolor{Gr}	0.9972&\cellcolor{Gr}	0.9534&\cellcolor{Gr}	0.8500&\cellcolor{Gr}	0.7448 \\
\hline
& ($+$) Drishti-GS  & 0.8253&	0.9995&	0.9966&	0.9691&	0.8915&	0.8042 \\ 
\textbf{RED-RCNN}&($+$) RIGA        &0.9138	&0.9997	&0.9994	&0.9268&	0.9203&	0.8523 \\ 
&($+$) REFUGE     &0.8785&	0.9998&	0.9993&	0.9614&	0.9181&	0.8486  \\ 
&(1460) &\cellcolor{Gr} 0.8725&\cellcolor{Gr}	0.9996&\cellcolor{Gr}	0.9984&\cellcolor{Gr}	0.9524&\cellcolor{Gr}	0.9099&\cellcolor{Gr}	0.8350 \\
\hline
& ($*$) Drishti-GS  & 0.8698	&0.9988	&0.9978	&0.8443	&0.8569	&0.7496 \\ 
\textbf{MRCNN}&($*$) RIGA        &0.8641	&0.9998	&0.9993	&0.9414	&0.9011 &	0.82\\ 
&($*$) REFUGE     &0.825&	0.9997&	0.9989&	0.922&	0.8708&	0.7712\\ 
 &(190) &\cellcolor{Gr} 0.8529&\cellcolor{Gr}	0.9994&\cellcolor{Gr}	0.9986&\cellcolor{Gr}	0.9025&\cellcolor{Gr}	0.8762&\cellcolor{Gr}	0.7802 \\
\hline
\textbf{RED-RCNN} & ($*$) Drishti-GS  &0.9855&	0.9965&	0.9964&	0.6794&	0.8043&	0.6727 \\ 
$+$: Train dataset &($*$) RIGA        &0.914&	0.9997&	0.9994&	0.925&	0.9194&	0.8509\\ 
$*$: Test dataset &($*$) REFUGE     &0.8438&	0.9999&	0.9992&	0.9656&	0.9006&	0.8192 \\ 
 \cellcolor{Gr} \textbf{Average}&(190) &\cellcolor{Gr} 0.9144&\cellcolor{Gr} 0.9987&\cellcolor{Gr} 0.9983&\cellcolor{Gr} 0.8567&\cellcolor{Gr} 0.8748 & \cellcolor{Gr}0.7809\\
\hline
\hline
\end{tabular}
\label{trte oc}
\end{table*}

\begin{table*}[h!]
\centering
\footnotesize
\caption{Performance comparison of MRCNN and RED-RCNN for OD segmentation on unseen data. The average performance indices are also provided. \textcolor{uptxt}{The unseen dataset has 1331 images comprised as follows: 50 (Drions-DB) + 81 (IDRiD) + 1200 (MESSIDOR).}}
\vspace{0.2cm}
\begin{tabular}{@{}|c|c|c|c|c|c|c|c|@{}}
\hline
\hline
\textbf{Network} & \textbf{Datasets} &\textbf{Sensitivity} & \textbf{Specificity} & \textbf{Accuracy} & \textbf{Precision} & \textbf{Dice} & \textbf{Jaccard} \\ 
&(Total no. of images)&& & & & & \\
\hline
&IDRiD     &0.9285 &0.9995  &0.9982 &0.9713 &0.9495 &0.9038  \\ 
\textbf{MRCNN}&Drions-DB     & 0.875& 0.9993  &0.9954 &0.9745 &0.9221 &0.8554 \\ 
&MESSIDOR     & 0.9329&	0.9997	&0.9991	&0.9735	&0.9528	&0.9098 \\
&(1331) &   \cellcolor{Gr}0.9121& \cellcolor{Gr} 0.9995 & \cellcolor{Gr} 0.9975 & \cellcolor{Gr} 0.9731 & \cellcolor{Gr} 0.9414 & \cellcolor{Gr} 0.8896\\
\hline
&IDRiD     &0.9241 &0.9992  &0.9979 &0.9562 &0.9399 &0.8866  \\ 
\textbf{RED-RCNN} & Drions-DB     &0.9547	&0.9986	&0.9973	&0.9572	&0.956	&0.9156\\
 & MESSIDOR    &0.9563&	0.9997	&0.9993&	0.9699	&0.9631	&0.9288\\
 \cellcolor{Gr} \textbf{Average} & (1331) & \cellcolor{Gr}0.945 &\cellcolor{Gr}0.9991 &\cellcolor{Gr}0.9981 &\cellcolor{Gr}0.9611 & \cellcolor{Gr} 0.953& \cellcolor{Gr}0.9103 \\
\hline
\hline
\end{tabular}
\label{tab genl}
\end{table*}

\begin{table*}[h!]
\centering
\footnotesize
\begin{center}
\caption{Performance comparison (on test dataset) of RED-RCNN (the proposed method) for optic disc segmentation with the state-of-the-art techniques. The symbol $-$ indicates that the corresponding value has not been reported by the authors. \textcolor{uptxt}{The unseen dataset used for evaluating RED-RCNN has 1521 images and is comprised as follows: 50 (Drions-DB) + 81 (IDRiD) + 1200 (MESSIDOR) + 10 (Drishti) + 81 (REFUGE) + 99 (RIGA).}}
\begin{tabular}{|c|c|c|c|c|c|c|c|}
\hline
\hline
\textbf{Algorithm} & \textbf{Datasets} &\textbf{Sensitivity} & \textbf{Specificity} & \textbf{Accuracy} & \textbf{Precision} & \textbf{Dice} & \textbf{Jaccard} \\ 
&(Total no. of images)&&&&&&\\
\hline
Dey et al. \cite{dey2019automatic} & IDRiD &0.923& 0.999  & --& --& 0.943& 0.896\\
& Dristhi-GS &0.932 &0.999  &-- &-- &0.958 &0.921\\
& RIM-ONE &0.953 & 0.961 &-- &-- &0.933 &0.88\\
& MESSIDOR &0.924 &0.976 &-- &-- &0.912 &0.857\\
& Drions-DB &0.919 &0.997 &-- &-- &0.913 &0.85\\
& (1583) &\cellcolor{Gr}0.9302 &\cellcolor{Gr}	0.9864	& & &\cellcolor{Gr} 0.931 &\cellcolor{Gr} \cellcolor{Gr}0.88\\
\hline
{Maninis et al. \cite{DRIU-2016}} &{Drions-DB} &{--} &{ -- } &{ -- }&{--}&{ 0.971} & \textcolor{uptxt}{0.944}\\
&{RIM-ONE} &{--} & {--}  & {--} &{--}& {0.959} & \textcolor{uptxt}{0.921}\\
&{(269)}&&&&&\cellcolor{Gr}{0.965}&\cellcolor{Gr}\textcolor{uptxt}{0.933}\\
\hline

\textcolor{uptxt}{Mohan et al. \cite{dm1}}&	\textcolor{uptxt}{MESSIDOR} &\textcolor{uptxt}{--}&	\textcolor{uptxt}{--}&	\textcolor{uptxt}{--}&	\textcolor{uptxt}{--}&\textcolor{uptxt}{0.957}&	\textcolor{uptxt}{0.92}\\
&	\textcolor{uptxt}{Drishti-GS}&	\textcolor{uptxt}{--}&\textcolor{uptxt}{	--}&\textcolor{uptxt}{	--}&	\textcolor{uptxt}{--}&\textcolor{uptxt}{	0.964}&	\textcolor{uptxt}{0.931}\\
&\textcolor{uptxt}{	Drions-DB}&	\textcolor{uptxt}{--}&\textcolor{uptxt}{	--}&\textcolor{uptxt}{	--}&\textcolor{uptxt}{	--}&\textcolor{uptxt}{	0.955}&	\textcolor{uptxt}{0.914}\\
& \textcolor{uptxt}{(1411)} &&&&&\cellcolor{Gr}\textcolor{uptxt}{0.959}&\cellcolor{Gr}\textcolor{uptxt}{0.922}\\
\hline
\textcolor{uptxt}{Mohan et al. \cite{dm2}}&	\textcolor{uptxt}{MESSIDOR} &\textcolor{uptxt}{--}&	\textcolor{uptxt}{--}&	\textcolor{uptxt}{--}&	\textcolor{uptxt}{--}&	\textcolor{uptxt}{0.968}&	\textcolor{uptxt}{0.939}\\
&	\textcolor{uptxt}{Drishti-GS}&	\textcolor{uptxt}{--}&\textcolor{uptxt}{	--}&\textcolor{uptxt}{	--}&	\textcolor{uptxt}{--}&\textcolor{uptxt}{	0.9713}&\textcolor{uptxt}{	0.947}\\
&\textcolor{uptxt}{	Drions-DB}&	\textcolor{uptxt}{--}&\textcolor{uptxt}{	--}&\textcolor{uptxt}{	--}&\textcolor{uptxt}{	--}&\textcolor{uptxt}{	0.966}&	\textcolor{uptxt}{0.935}\\
& \textcolor{uptxt}{(1411)} &&&&&\cellcolor{Gr}\textcolor{uptxt}{0.968}&\cellcolor{Gr}\textcolor{uptxt}{0.94}\\
\hline
Aretm et al. \cite{sevastopolsky2017optic}& Drions-DB &-- &--  &-- &-- &0.94 &0.89\\
& RIM-ONE  &-- &--  &-- &-- &0.95 &0.89\\
& (269) & & &  & &\cellcolor{Gr}0.945 & \cellcolor{Gr}0.89\\
\hline
Zilly et al. \cite{zilly2017glaucoma}& Drishti-GS &-- &-- &-- &--  &0.973 & \textcolor{uptxt}{0.947}\\
 & RIM-ONE  &-- &-- &-- &--  &0.942 &\textcolor{uptxt}{0.89}\\
 & \textcolor{uptxt}{MESSIDOR}  &-- &-- &-- &--  &\textcolor{uptxt}{0.90} &\textcolor{uptxt}{0.818}\\
& (1409) & & & &  &\cellcolor{Gr}\textcolor{uptxt}{0.938} & \cellcolor{Gr}\textcolor{uptxt}{0.885}\\
\hline
Singh et al. \cite{singh2018refuge}& REFUGE &-- &-- &0.934 &-- &-- &--\\
&(400)&&&&&&\\
\hline
Edupuganti et al. \cite{edupuganti2018automatic} & Drishti-GS &-- &-- &-- &-- &0.967&0.936\\
& (51) &&&&&&\\
\hline
Agrawal et al. \cite{agrawal2018enhanced} & REFUGE&-- &-- &-- &-- &0.88&0.786\\
& (400) &&&&&&\\
\hline
\textcolor{uptxt}{Fu et al. \cite{fu2018joint}} & \textcolor{uptxt}{ORIGA} &\textcolor{uptxt}{--}&\textcolor{uptxt}{--}&\textcolor{uptxt}{--}&\textcolor{uptxt}{--}&\textcolor{uptxt}{0.963}&\textcolor{uptxt}{0.926}\\
& \textcolor{uptxt}{(325)} &&&&&&\\
\hline
\textcolor{uptxt}{Al-Bander et al. \cite{al2018dense}} &\textcolor{uptxt}{ORIGA}&\textcolor{uptxt}{0.961}&	\textcolor{uptxt}{0.999}&	\textcolor{uptxt}{0.999}&\textcolor{uptxt}{	--}&\textcolor{uptxt}{	0.9653}&	\textcolor{uptxt}{0.9334}\\
&\textcolor{uptxt}{Drions-DB}&\textcolor{uptxt}{	0.9232}&\textcolor{uptxt}{	0.999}&\textcolor{uptxt}{  0.997}&\textcolor{uptxt}{	--}&\textcolor{uptxt}{	0.9415}&\textcolor{uptxt}{	0.8912}\\
&\textcolor{uptxt}{ONHSD}&	\textcolor{uptxt}{0.9376}&	\textcolor{uptxt}{0.999}&\textcolor{uptxt}{	0.999}&\textcolor{uptxt}{	--}&\textcolor{uptxt}{	0.9556}&\textcolor{uptxt}{	0.9155}\\
&\textcolor{uptxt}{Drishti-GS}&	\textcolor{uptxt}{0.9268}&	\textcolor{uptxt}{0.999}&\textcolor{uptxt}{	0.997}&	\textcolor{uptxt}{--}&\textcolor{uptxt}{	0.949}&	\textcolor{uptxt}{0.9042}\\
&\textcolor{uptxt}{RIM-ONE}&\textcolor{uptxt}{	0.8737}&\textcolor{uptxt}{	0.998}&	\textcolor{uptxt}{0.992}&	\textcolor{uptxt}{--}&	\textcolor{uptxt}{0.9036}&\textcolor{uptxt}{0.8289}\\
& \textcolor{uptxt}{(664)} &\cellcolor{Gr}\textcolor{uptxt}{0.924}&\cellcolor{Gr}\textcolor{uptxt}{0.999}&\cellcolor{Gr}\textcolor{uptxt}{0.997}&&\cellcolor{Gr}\textcolor{uptxt}{0.943}&\cellcolor{Gr}\textcolor{uptxt}{0.895}\\
\hline
\textcolor{uptxt}{Liu et al. \cite{liu2019ddnet}} & \textcolor{uptxt}{ORIGA} &\textcolor{uptxt}{--}&\textcolor{uptxt}{--}&\textcolor{uptxt}{--}&\textcolor{uptxt}{--}&\textcolor{uptxt}{0.972}&\textcolor{uptxt}{0.946}\\
& \textcolor{uptxt}{(325)} &&&&&&\\
\hline
\textcolor{uptxt}{Jiang et al. \cite{jiang2019jointrcnn}} & \textcolor{uptxt}{ORIGA} &\textcolor{uptxt}{--}&\textcolor{uptxt}{--}&\textcolor{uptxt}{--}&\textcolor{uptxt}{--}&\textcolor{uptxt}{0.967}&\textcolor{uptxt}{0.937}\\
& \textcolor{uptxt}{(325)} &&&&&&\\
\hline
Yu et al. \cite{yu2019robust}& RIGA (Magrabia) &--  &-- &-- &-- & 0.969 & 0.942\\
& Drishti-GS &--  &-- &-- &-- &0.964 &0.932\\
& RIM-ONE &-- &--  &-- &-- &0.949 &0.907\\
&(324)&&&&&\cellcolor{Gr}0.96&\cellcolor{Gr}0.927\\
\hline
\textcolor{uptxt}{Yin et al. \cite{yin2019pm}}&	\textcolor{uptxt}{ORIGA}&\textcolor{uptxt}{	--}&\textcolor{uptxt}{	--}&\textcolor{uptxt}{	--}&\textcolor{uptxt}{	--}&\textcolor{uptxt}{	0.966}&\textcolor{uptxt}{	0.934}\\
&\textcolor{uptxt}{	REFUGE}&\textcolor{uptxt}{	--}&\textcolor{uptxt}{	--}&\textcolor{uptxt}{	--}&\textcolor{uptxt}{	--}&\textcolor{uptxt}{	0.954}&\textcolor{uptxt}{	0.912}\\
& \textcolor{uptxt}{(725)} &&&&&\cellcolor{Gr}\textcolor{uptxt}{0.96}&\cellcolor{Gr}\textcolor{uptxt}{0.923}\\
\hline
Kumar et al. \cite{RDR} &Drishti-GS &0.9962 & 0.9698  &0.9732&--&0.9282&0.8707\\
&Drions-DB &0.8410 & 0.996  &0.985&--& 0.9062 & 0.8341\\
&MESSIDOR &0.8840 & 0.9894 &0.9882&--& 0.8760 & 0.7858\\
&Forus &0.9457 & 0.9957  & 0.9911 &--& 0.9642 & 0.9319\\
&Bosch &0.9649 & 0.9998  & 0.9968 &--& 0.9531 & 0.9109\\
&(1597)&\cellcolor{Gr}0.9263&\cellcolor{Gr}0.9901&\cellcolor{Gr}0.9868&&\cellcolor{Gr}0.9255&\cellcolor{Gr}0.8666\\
\hline
Kim et al. \cite{kim2019optic} & MESSIDOR & 0.98 & 0.99  &0.99 &--&\textcolor{uptxt}{0.969}&0.940\\
& (1200) &&&&&&\\
\hline
\textbf{Proposed method}
& Drishti-GS  &0.9839&	0.9975&	0.9971&	0.9216&	0.9517&	0.9078 \\
& RIGA        &0.9623 &	0.9994	&0.9989	 &0.9615 &	0.9619 &	0.9266\\ 
& REFUGE     &0.9572 &	0.9996 &	0.9989 &	0.9743 &	0.9657 &	0.9336 \\
\textbf{RED-RCNN}& IDRiD     &0.9241 &0.9992  &0.9979 &0.9562 &0.9399 &0.8866  \\ 
 &Drions-DB     &0.9547	&0.9986	&0.9973	&0.9572	&0.956	&0.9156\\
 & MESSIDOR     &0.9563&	0.9997	&0.9993&	0.9699	&0.9631	&0.9288\\
\cellcolor{Gr} \textbf{Average} & (1521) & \cellcolor{Gr}0.9564 & \cellcolor{Gr}0.999&\cellcolor{Gr} 0.9982& \cellcolor{Gr}0.9568&\cellcolor{Gr} 0.9564& \cellcolor{Gr}0.9165\\
\hline
\hline
\end{tabular}
\label{comp_od}
\end{center}
\end{table*}

\begin{table*}[h!]
\centering
\footnotesize
\vspace{0.5cm}
\caption{Performance comparison (on test dataset) of RED-RCNN for optic cup segmentation with the state-of-the-art techniques. A $-$ indicates that the corresponding value has not been reported by the authors. \textcolor{uptxt}{The unseen dataset used for evaluating RED-RCNN has 190 images and is comprised as follows: 10 (Drishti) + 81 (REFUGE) + 99 (RIGA).}}
\begin{tabular}{|c|c|c|c|c|c|c|c|}
\hline
\hline
\textbf{Algorithm} & \textbf{Datasets} &\textbf{Sensitivity} & \textbf{Specificity} & \textbf{Accuracy} & \textbf{Precision} & \textbf{Dice} & \textbf{Jaccard} \\ 
&(Total no. of images used)&&&&&&\\
\hline
Yang et al. \cite{yang2018efficient} &RIM-ONE & -- & --  &-- &--&0.813&0.685\\
& (94) &&&&&&\\
\hline
Aretm et al. \cite{sevastopolsky2017optic}& Drions-DB &-- &--  &-- &-- &0.85 &0.75\\
& RIM-ONE  &-- &--  &-- &-- &0.82 &0.69\\
& (269) & & &  & &\cellcolor{Gr}0.835 &\cellcolor{Gr}0.72\\
\hline
Zilly et al. \cite{zilly2017glaucoma}& Drishti-GS &-- &-- &-- &--  &0.871 &0.771\\
 & RIM-ONE  &-- &-- &-- &--  &0.824 &0.701\\
& (209) & & & &  &\cellcolor{Gr}0.847 &\cellcolor{Gr}0.736\\
\hline
Singh et al. \cite{singh2018refuge}& REFUGE &-- &-- &0.8341 &-- &-- &--\\
&(400)&&&&&&\\
\hline
Edupuganti et al. \cite{edupuganti2018automatic} & Drishti-GS &-- &-- &-- &-- &0.897&0.813\\
& (51) &&&&&&\\
\hline
Agrawal et al. \cite{agrawal2018enhanced} & REFUGE&-- &-- &-- &-- &0.64&0.471\\
& (400) &&&&&&\\
\hline
\textcolor{uptxt}{Fu et al. \cite{fu2018joint}} & \textcolor{uptxt}{ORIGA} &\textcolor{uptxt}{--}&\textcolor{uptxt}{--}&\textcolor{uptxt}{--}&\textcolor{uptxt}{--}&\textcolor{uptxt}{0.87}&\textcolor{uptxt}{0.77}\\
& \textcolor{uptxt}{(325)} &&&&&&\\
\hline
\textcolor{uptxt}{Al-Bander et al. \cite{al2018dense}}&	\textcolor{uptxt}{ORIGA}&	\textcolor{uptxt}{0.9195}&	\textcolor{uptxt}{0.999}&	\textcolor{uptxt}{0.998}&\textcolor{uptxt}{	--}&	\textcolor{uptxt}{0.8659}&	\textcolor{uptxt}{0.7688}\\
&	\textcolor{uptxt}{Drishti-GS}&	\textcolor{uptxt}{0.7413}&\textcolor{uptxt}{	0.999}&	\textcolor{uptxt}{0.995}&\textcolor{uptxt}{	--}&\textcolor{uptxt}{	0.8282}&\textcolor{uptxt}{	0.7113}\\
&\textcolor{uptxt}{	RIM-ONE}&\textcolor{uptxt}{	0.9052}&\textcolor{uptxt}{	0.994}&\textcolor{uptxt}{	0.993}&\textcolor{uptxt}{	--}&\textcolor{uptxt}{	0.6903}&\textcolor{uptxt}{	0.5567}\\
& \textcolor{uptxt}{(455)} &\cellcolor{Gr}\textcolor{uptxt}{0.855}&\cellcolor{Gr}\textcolor{uptxt}{0.997}&\cellcolor{Gr}\textcolor{uptxt}{0.995}&&\cellcolor{Gr}\textcolor{uptxt}{0.795}&\cellcolor{Gr}\textcolor{uptxt}{0.679}\\
\hline
\textcolor{uptxt}{Liu et al. \cite{liu2019ddnet} }& \textcolor{uptxt}{ORIGA} &\textcolor{uptxt}{--}&\textcolor{uptxt}{--}&\textcolor{uptxt}{--}&\textcolor{uptxt}{--}&\textcolor{uptxt}{0.886}&\textcolor{uptxt}{0.796}\\
& \textcolor{uptxt}{(325)} &&&&&&\\
\hline
\textcolor{uptxt}{Jiang et al. \cite{jiang2019jointrcnn}} &\textcolor{uptxt}{ORIGA} &\textcolor{uptxt}{--}&\textcolor{uptxt}{--}&\textcolor{uptxt}{--}&\textcolor{uptxt}{--}&\textcolor{uptxt}{0.883}&\textcolor{uptxt}{0.791}\\
& \textcolor{uptxt}{(325)} &&&&&&\\
\hline
Yu et al. \cite{yu2019robust}& RIGA (Magrabia) &--  &-- &-- &-- &0.891 &0.812\\
& Drishti-GS &--  &-- &-- &-- &\textcolor{uptxt}{0.874} &0.781\\
& RIM-ONE &-- &--  &-- &-- &0.793 &0.683\\
&(324)&&&&&\cellcolor{Gr}\textcolor{uptxt}{0.852}&\cellcolor{Gr}\textcolor{uptxt}{0.758}\\
\hline
\textcolor{uptxt}{ Yin et al. \cite{yin2019pm}}&	\textcolor{uptxt}{ORIGA}&	\textcolor{uptxt}{--}&	\textcolor{uptxt}{--}&\textcolor{uptxt}{	--}&\textcolor{uptxt}{	--}&\textcolor{uptxt}{	0.884}&	\textcolor{uptxt}{0.792}\\
&\textcolor{uptxt}{	REFUGE}&\textcolor{uptxt}{	--}&\textcolor{uptxt}{	--}&\textcolor{uptxt}{	--}&\textcolor{uptxt}{	--}&\textcolor{uptxt}{	0.875}&\textcolor{uptxt}{	0.777}\\
& \textcolor{uptxt}{(725)} &&&&&\cellcolor{Gr}\textcolor{uptxt}{0.88}&\cellcolor{Gr}\textcolor{uptxt}{0.785}\\
\hline
Kumar et al. \cite{RDR} &Drishti-GS &0.8589 & 0.9698  &0.9690&--&0.7532&0.6143\\
&Drions-DB &0.6175 & 0.9979  &0.9884&--& 0.73 & 0.5898\\
&MESSIDOR &0.8357 & 0.9883 &0.9823&--& 0.7303 & 0.5911\\
&Forus &0.7466 & 0.9952  & 0.9867 &--& 0.8027 & 0.6801\\
&Bosch &0.7584 & 0.9959  & 0.9917 &--& 0.7213 & 0.5804\\
&(436)&\cellcolor{Gr}0.7634&\cellcolor{Gr}0.9894&\cellcolor{Gr}0.9836&&\cellcolor{Gr}0.742&\cellcolor{Gr}0.611\\
\hline
\textbf{Proposed method}& Drishti-GS &0.9855&	0.9965&	0.9964&	0.6794&	0.8043&	0.6727\\
\textbf{RED-RCNN}& RIGA      &0.914&	0.9997&	0.9994&	0.925&	0.9194&	0.8509 \\
& REFUGE    &0.8438&	0.9999&	0.9992&	0.9656&	0.9006&	0.8192  \\ 
\cellcolor{Gr} \textbf{Average}& \textcolor{uptxt}{(190)} &\cellcolor{Gr} 0.9144&\cellcolor{Gr} 0.9987&\cellcolor{Gr} 0.9983&\cellcolor{Gr} 0.8567&\cellcolor{Gr} 0.8748 & \cellcolor{Gr}0.7809\\
\hline
\hline
\end{tabular}
\label{comp_oc}
\end{table*}

\subsection{Performance Measures}
\indent The standard performance measures used for OD and OC segmentation are Sensitivity, Specificity, Accuracy, Precision, Dice index, and Jaccard index \cite{dice,jaccard}. Sensitivity measures the magnitude of positives that are correctly identified as positives, whereas Specificity measures the magnitude of negatives that are correctly identified as negatives. Precision expresses the proportion of all predictions that are actually relevant or present in the ground-truth. Accuracy is the ratio between the correct predictions and the total number of predictions.\\
\indent The Dice and Jaccard indices using network predictions (A) and the ground truth (B) are: $\text{Jaccard} = \frac{|A\;\cap\;B|}{|A\;\cup\;B|},\text{and } \text{Dice} =\frac{2\,|A\;\cap\;B|}{|A|\;+\;|B|}.$
We compute the Pearson correlation coefficient (PCC) between the CDRs obtained from the ground truth and the prediction.

\begin{figure*}[h!]
    \centering
    $\begin{array}{ccccc}
    \includegraphics[width=2.4cm, height = 1.8cm]{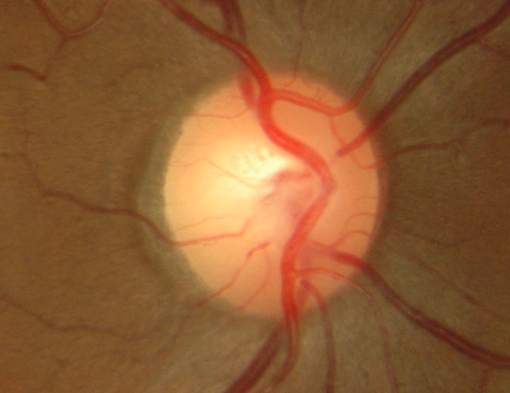} &
    \includegraphics[width=2.4cm, height = 1.8cm]{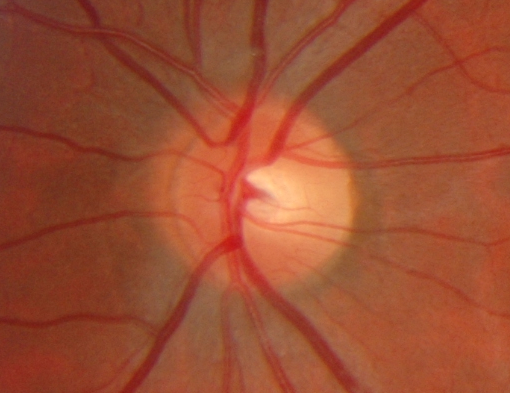} &
    \includegraphics[width=2.4cm, height = 1.8cm]{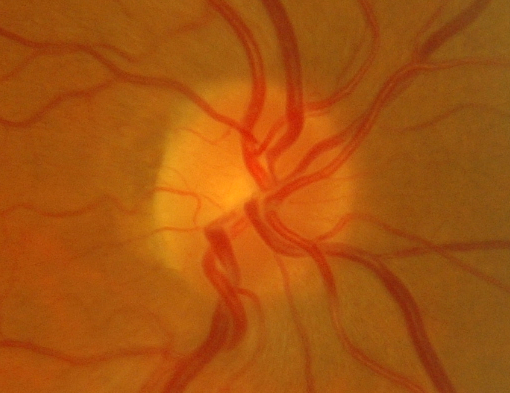} &
    \includegraphics[width=2.4cm, height = 1.8cm]{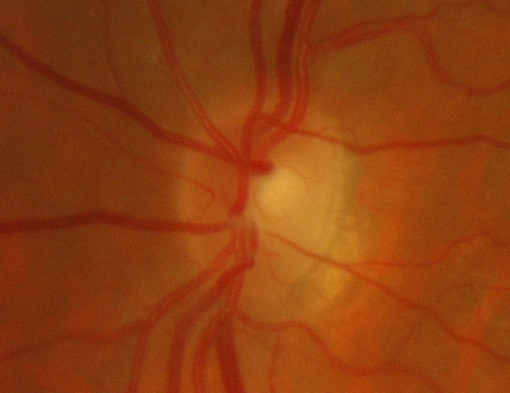} &
    \includegraphics[width=2.4cm, height = 1.8cm]{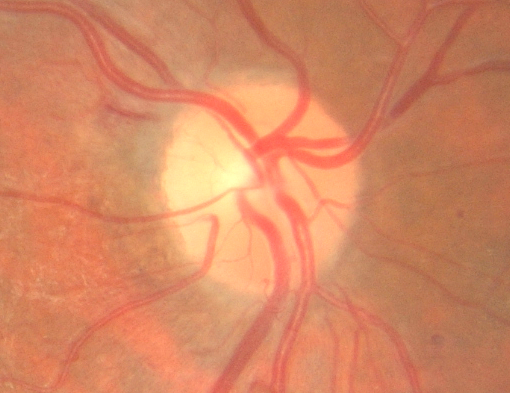}\\
    \mbox{(a1)}&\mbox{(a2)}&\mbox{(a3)}&\mbox{(a4)}&\mbox{(a5)}\\
     \includegraphics[width=2.4cm, height = 1.8cm]{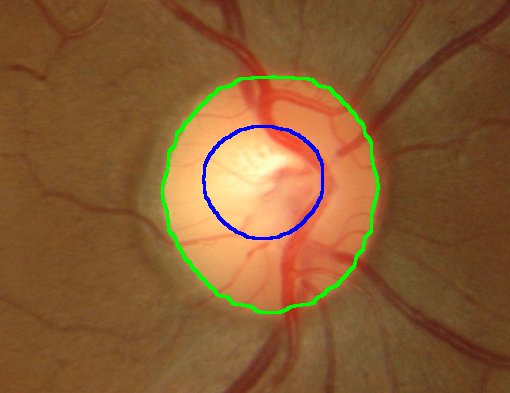} &
    \includegraphics[width=2.4cm, height = 1.8cm]{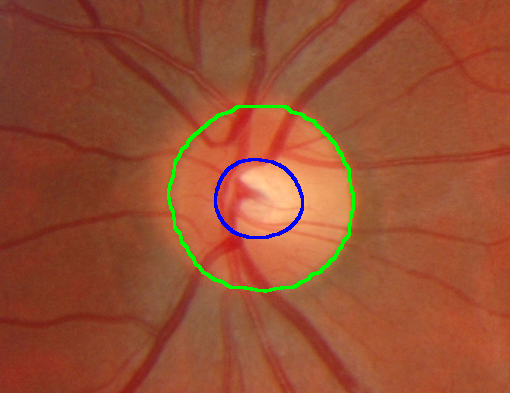} &
    \includegraphics[width=2.4cm, height = 1.8cm]{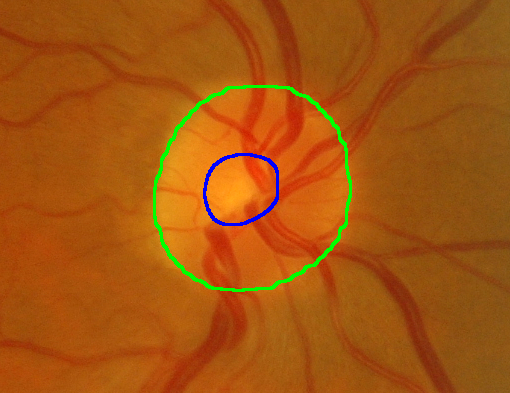} &
    \includegraphics[width=2.4cm, height = 1.8cm]{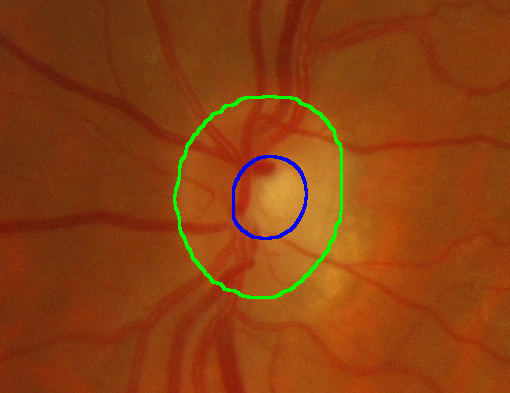} &
    \includegraphics[width=2.4cm, height = 1.8cm]{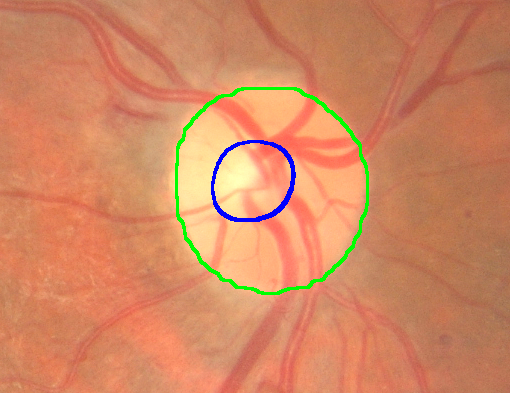}\\
    \mbox{(b1)}&\mbox{(b2)}&\mbox{(b3)}&\mbox{(b4)}&\mbox{(b5)}\\
     \includegraphics[width=2.4cm, height = 1.8cm]{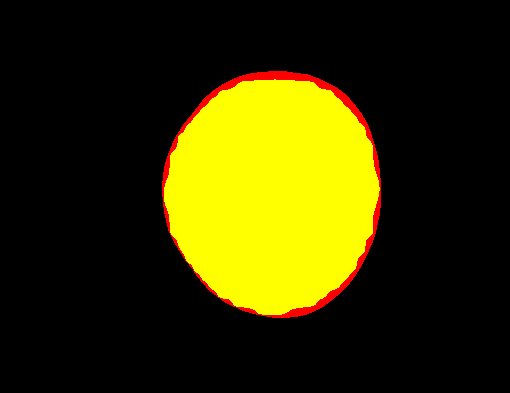} &
     \includegraphics[width=2.4cm, height = 1.8cm]{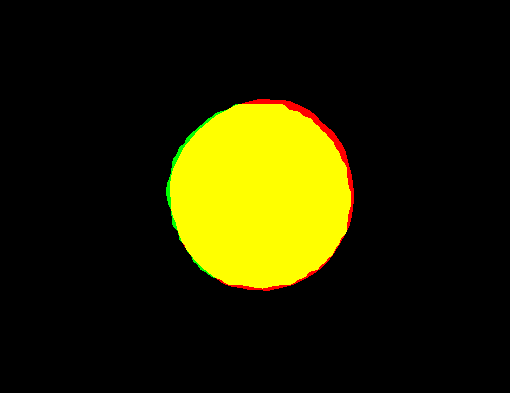} &
    \includegraphics[width=2.4cm, height = 1.8cm]{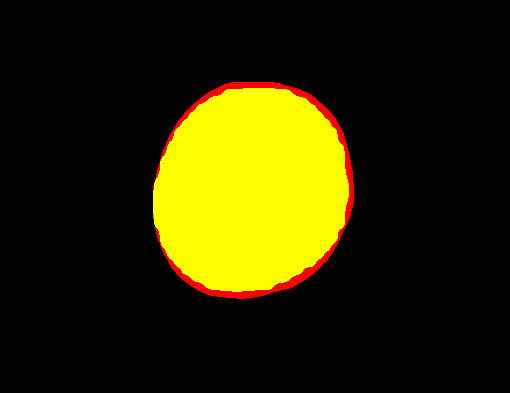} &
    \includegraphics[width=2.4cm, height = 1.8cm]{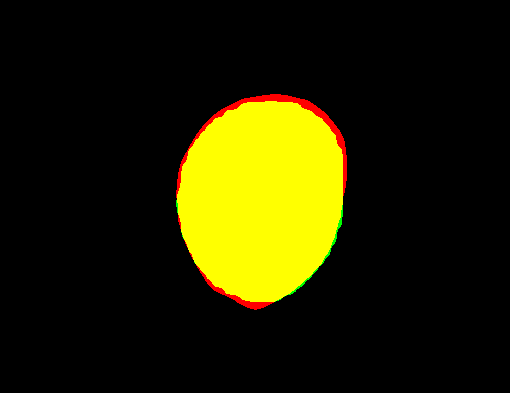} &
    \includegraphics[width=2.4cm, height = 1.8cm]{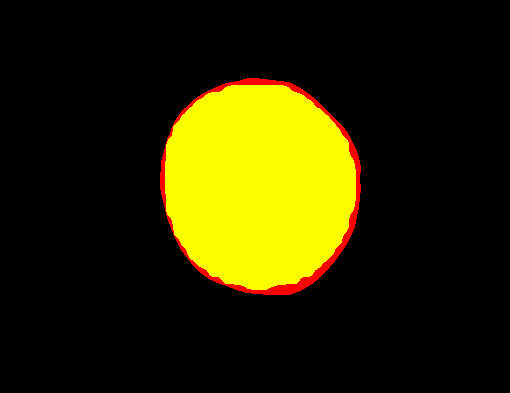}\\
    \mbox{(c1)}&\mbox{(c2)}&\mbox{(c3)}&\mbox{(c4)}&\mbox{(c5)}\\
    \includegraphics[width=2.4cm, height = 1.8cm]{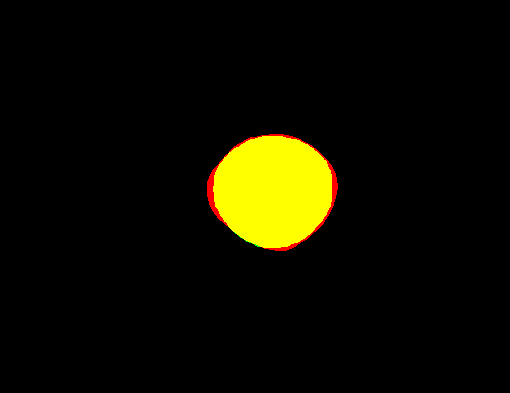} &
    \includegraphics[width=2.4cm, height = 1.8cm]{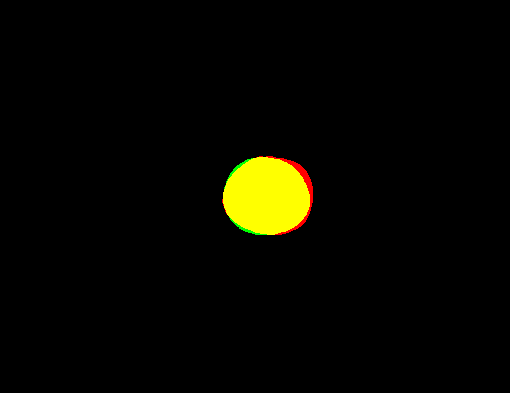} &
    \includegraphics[width=2.4cm, height = 1.8cm]{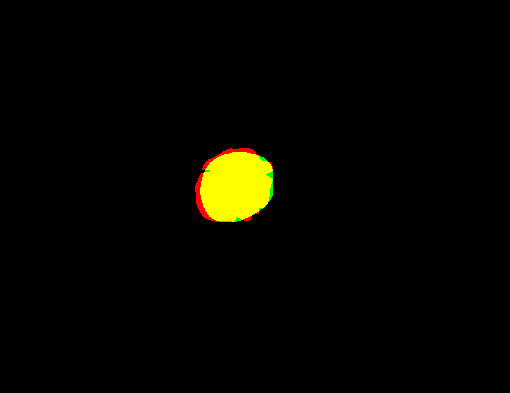} &
    \includegraphics[width=2.4cm, height = 1.8cm]{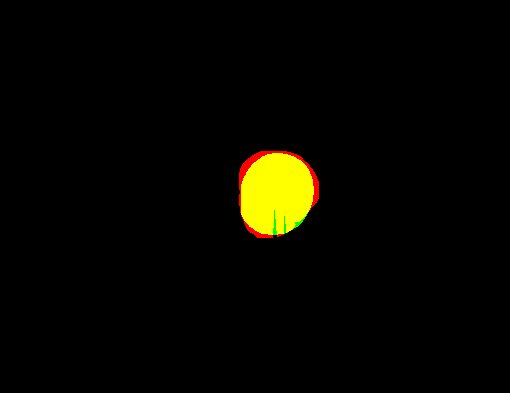} &
    \includegraphics[width=2.4cm, height = 1.8cm]{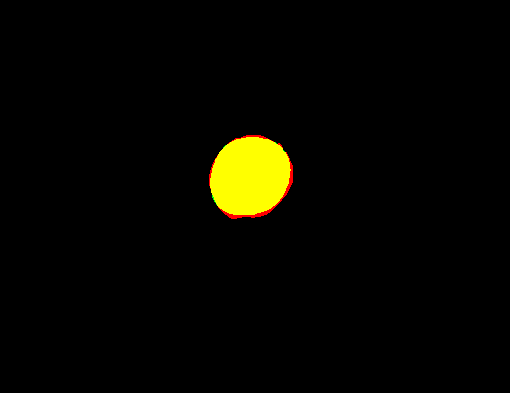}\\
        \mbox{(d1)}&\mbox{(d2)}&\mbox{(d3)}&\mbox{(d4)}&\mbox{(d5)}\\
    \end{array}$
    \caption{\protect\rgbsymbol\, Segmentation of optic disc and optic cup using RED-RCNN. (a1)$-$(a5) are fundus images; (b1)$-$(b5) show the OD outline (green) \& OC outline (blue), respectively ; (c1)$-$(c5) show the OD segmentation performance; and (d1)$-$(d5) show the OC segmentation performance; yellow: true positive; red: false negative; green: false positive.}
    \label{segres}
\end{figure*} \vspace{-0.2cm}

\begin{table}[h]
\scriptsize
\caption{\textcolor{uptxt}{Database-specific comparison of simultaneous OD and OC segmentation techniques.}}
\begin{tabular*}{0.48\textwidth}{|c|c|c|c|c|c|}
\hline
\hline
\multirow{2}{*}{\textbf{Dataset}}  & \multirow{2}{*}{\textbf{Algorithm}} & \multicolumn{2}{c|}{\textbf{OD}} & \multicolumn{2}{c|}{\textbf{OC}}  \\ \cline{3-6}
                      &                           & \textbf{Dice} & \textbf{Jaccard}            & \textbf{Dice} & \textbf{Jaccard}             \\
\hline
            & Zilly et al. \cite{zilly2017glaucoma}
                & 0.942 & 0.89 & \textbf{0.842} & \textbf{0.701} \\
RIM-ONE     & Al-Bander et al. \cite{al2018dense}
                & 0.903 & 0.828 & 0.69 & 0.556 \\
            & Yu et al. \cite{yu2019robust}
                & \textbf{0.949} & \textbf{0.907} & 0.793 & 0.683 \\
\hline
            & Fu et al. \cite{fu2018joint}
                & 0.963 & 0.926 & 0.87 & 0.77 \\
            & Al-Bander et al. \cite{al2018dense}
                & 0.965 & 0.933 & 0.865 & 0.768 \\
ORIGA      & Liu et al. \cite{liu2019ddnet}
                & \textbf{0.972} & \textbf{0.946} & \textbf{0.886} & \textbf{0.796} \\
            & Jiang et al. \cite{jiang2019jointrcnn}
                & 0.967 & 0.937 & 0.883 & 0.791 \\
            & Yin et al. \cite{yin2019pm}
                & 0.966 & 0.934 & 0.884 & 0.792 \\
\hline
            & Zilly et al. \cite{zilly2017glaucoma}
                & \textbf{0.973} & \textbf{0.947} & 0.871 & 0.771 \\
            & Edupuganti et al. \cite{edupuganti2018automatic}
                & 0.967 & 0.936 & \textbf{0.897} & \textbf{0.813} \\
Drishti-GS  & Al-Bander et al. \cite{al2018dense}
                & 0.949 & 0.904 & 0.828 & 0.711 \\
            & Yu et al. \cite{yu2019robust}
                & 0.964 & 0.932 & 0.874 & 0.781 \\
            & Kumar et al. \cite{RDR}
                & 0.928 & 0.87 & 0.75 & 0.61 \\
            & \textbf{Proposed method}
                & 0.951 & 0.907 & 0.804 & 0.672 \\
\hline
RIGA       & Yu et al. \cite{yu2019robust}
                & \textbf{0.969} & \textbf{0.942} & 0.891 & 0.812 \\
            & \textbf{Proposed method}
                & 0.961 & 0.926 & \textbf{0.914} & \textbf{0.851} \\
\hline
REFUGE     & Agrawal et al. \cite{agrawal2018enhanced}
                & 0.88 & 0.786 & 0.64 & 0.471 \\
            & Yin et al. \cite{yin2019pm}
                & 0.954 & 0.912 & 0.875 & 0.777 \\
            & \textbf{Proposed method}
                & \textbf{0.965} & \textbf{0.933} & \textbf{0.90} & \textbf{0.819} \\
            
\hline
\hline
\end{tabular*}
\label{tab:dtst} 
\end{table}

\subsection{Results} 
\begin{figure}[t!]
    \centering
    $\begin{array}{cc}
    \hspace{-0.2cm}
     \includegraphics[width=0.25\textwidth]{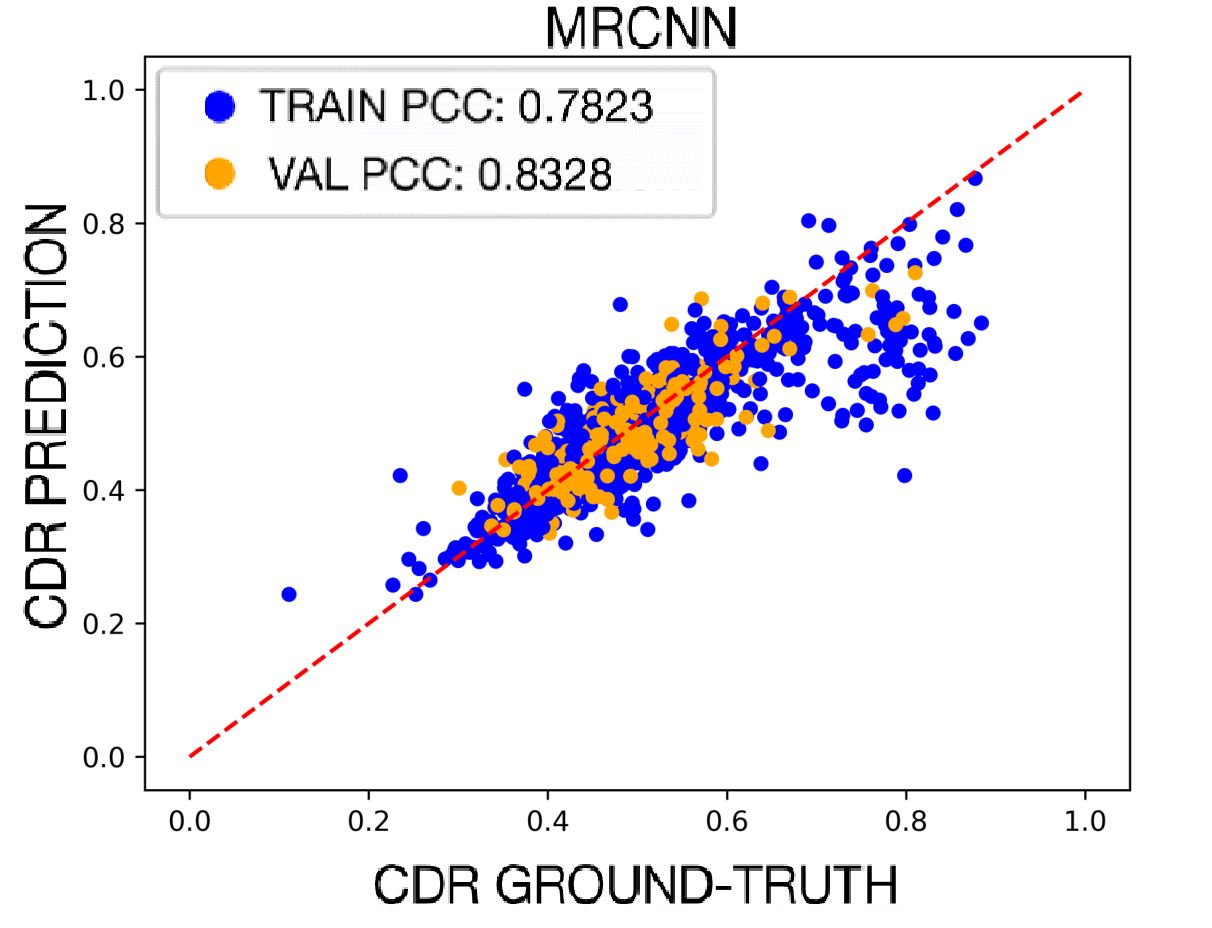} &
    \hspace{-0.4cm} \includegraphics[width=0.25\textwidth]{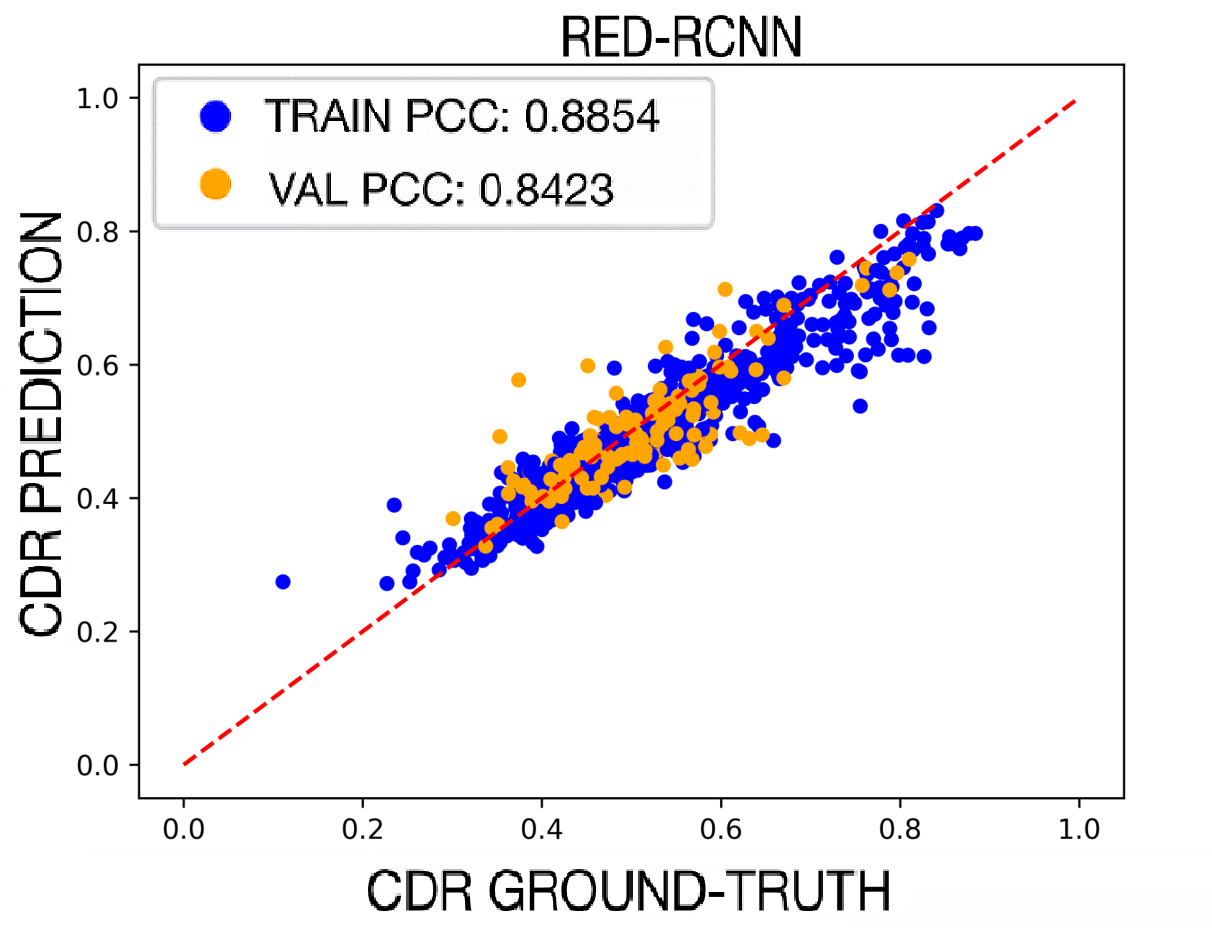}\\
     \mbox{(a)} & \mbox{(b)}\\
    \end{array}$
    \caption{Pearson's correlation coefficient (PCC) between the vertical CDRs computed from the ground-truth and the prediction obtained using (a) MRCNN, and (b) RED-RCNN for both training and validation data. The PCC values are higher for RED-RCNN than MRCNN.}
  \label{pccft1}
\end{figure}

\indent {\it MRCNN versus RED-RCNN}: The first set of comparisons pertains to the OD segmentation performance. The results of both training and test data performance are presented in Table~\ref{trte od}.
The training data performance corresponds to the scenario where the entire dataset was used for training and the validation was also carried out on the entire dataset. The test data performance corresponds to the scenario where 90\% of the dataset (with a 70\%$-$20\% split as explained in Sec.~\ref{net_tr}) was used for training, and the results have been reported on the unseen 10\%. We observe that the training data performance of both the models is comparable for most performance measures with RED-RCNN being marginally better than MRCNN in terms of Sensitivity and Dice and Jaccard indices. On the other hand, the test data performance of RED-RCNN is clearly superior to that of MRCNN. A similar set of results for OC segmentation is provided in Table~\ref{trte oc}. Overall, we observe that RED-RCNN is superior to MRCNN on both training and test datasets. Proceeding further, we validated the performance of MRCNN and RED-RCNN on completely new datasets (IDRiD, Drions-DB, MESSIDOR), i.e., datasets not used even partially for training. The OD segmentation results are shown in Table~\ref{tab genl}. A similar table could not be generated for OC segmentation since these datasets do not have ground-truth OC annotations (cf. Table~\ref{datasets}). The results show that RED-RCNN is superior to MRCNN and is better at generalizing to unseen data. Hence, for the subsequent comparisons with respect to the state-of-the-art techniques, we use RED-RCNN \textcolor{uptxt}{and show its OD and OC segmentation performance in Fig.~\ref{segres}.}\\
\begin{figure}[h!]
  \centering
  $\begin{array}{ccc}
    \includegraphics[width=2.4cm, height = 1.8cm]{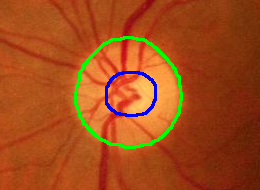} &
    \includegraphics[width=2.4cm, height = 1.8cm]{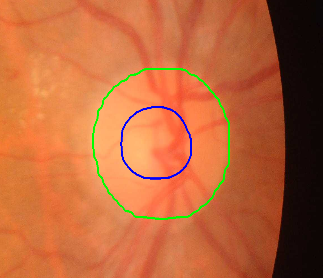} &
    \includegraphics[width=2.4cm, height = 1.8cm]{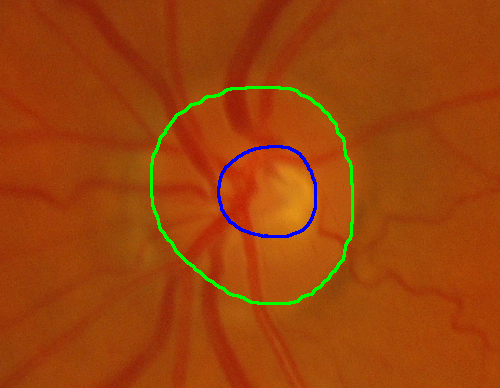} \\
    \mbox{(a1)}&\mbox{(b1)}&\mbox{(c1)}\\
    \includegraphics[width=2.4cm, height = 1.8cm]{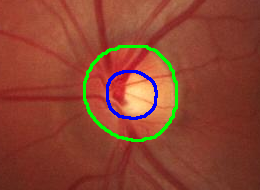} &
    \includegraphics[width=2.4cm, height = 1.8cm]{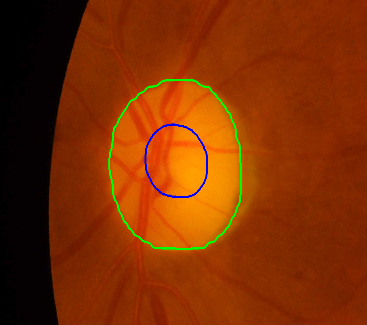} &
    \includegraphics[width=2.4cm, height = 1.8cm]{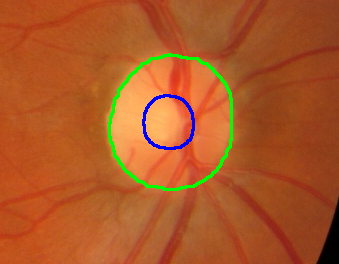} \\
    \mbox{(a2)}&\mbox{(b2)}&\mbox{(c2)}\\
    \includegraphics[width=2.4cm, height = 1.8cm]{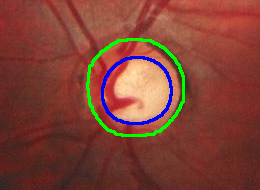} &
    \includegraphics[width=2.4cm, height = 1.8cm]{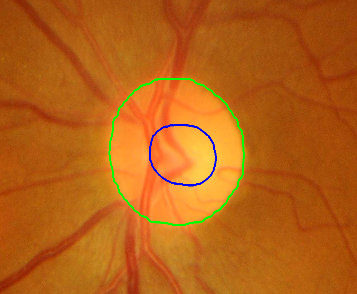} &
    \includegraphics[width=2.4cm, height = 1.8cm]{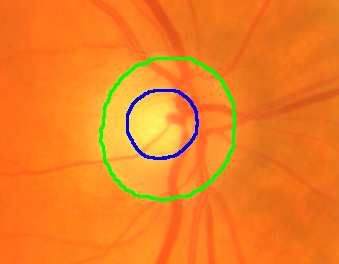} \\
    \mbox{(a3)}&\mbox{(b3)}&\mbox{(c3)}\\
    \includegraphics[width=2.4cm, height = 1.8cm]{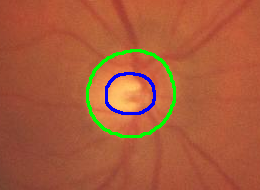} &
    \includegraphics[width=2.4cm, height = 1.8cm]{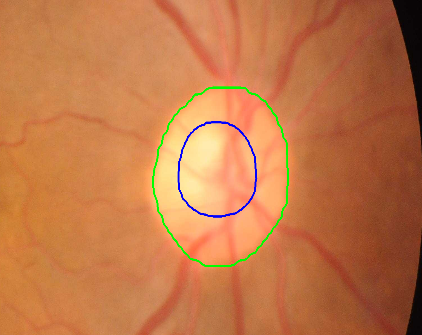} &
    \includegraphics[width=2.4cm, height = 1.8cm]{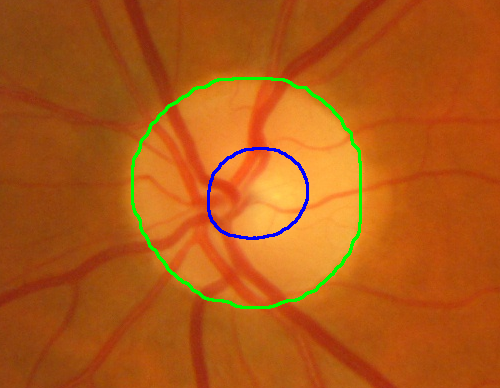} \\
    \mbox{(a4)}&\mbox{(b4)}&\mbox{(c4)}\\
    \end{array}$
    \caption{\protect\rgbsymbol\, \textcolor{uptxt}{Segmentation performance on unseen data (green: OD outline; blue: OC outline). The images (a1)$-$(a4) are taken from Drions-DB \cite{carmona2008identification}; (b1)$-$(b4) from IDRiD \cite{idrid}; and (c1)$-$(c4) from MESSIDOR \cite{decenciere_feedback_2014}.}}
    \label{UnseenOC}
    \end{figure}
\indent {\it RED-RCNN versus the state-of-the-art techniques}:
We compare the test data performance of the proposed method against the state-of-the-art techniques for OD (Table \ref{comp_od}) and OC (Table \ref{comp_oc}) segmentation. The results indicate that the overall performance of the proposed model is significantly better than most state-of-the-art approaches. \textcolor{uptxt}{The database-specific comparison of simultaneous OD and OC segmentation techniques is shown in Table~\ref{tab:dtst}. We observe a small decrease in the segmentation performance on the Drishti-GS \cite{sivaswamy2015comprehensive} dataset. The images in the Drishti-GS dataset have poor contrast in comparison with those in the RIGA and REFUGE datasets used for training. Drishti-GS has 101 images, whereas RIGA and REFUGE have 750 and 800 images, respectively. The mismatch in contrast accompanied by the differences in the dataset sizes could have been the reason for poor performance on this specific dataset. Examples illustrating segmentation performance of the proposed method on unseen data taken from \cite{idrid, carmona2008identification, decenciere_feedback_2014} are shown in Fig.~\ref{UnseenOC}}. The Pearson correlation coefficient computed between the CDR predictions and the CDR ground-truth is higher for RED-RCNN than for MRCNN as shown in Fig.~\ref{pccft1}.\\
\indent {\it Two-stage classification of Glaucoma severity}: \textcolor{uptxt}{Based on the OD and OC  boundaries generated by MRCNN and RED-RCNN, we compute the vertical CDR from the vertical OD and OC diameters and perform a two-stage classification of glaucoma severity. The threshold on the vertical CDR for two-stage classification is set to 0.5, in accordance with the International Council of Ophthalmology (ICO) guidelines \cite{ico}.} The validations are carried out on the REFUGE dataset because, unlike the other datasets, it not only has OD and OC outlines, but also the ground-truth for two-stage glaucoma severity grading, which is necessary to compute the Sensitivity, Specificity, and Overall Classification Accuracy (OCA). The results are shown in Table~\ref{tab:cdrd}. While the Sensitivity of MRCNN is higher than that of RED-RCNN, the Specificity and OCA of RED-RCNN are significantly better than that of MRCNN. The superior segmentation accuracy of RED-RCNN translates to a superior accuracy in glaucoma severity grading.

\section{Conclusions}
\noindent We addressed the problem of optic disc and cup segmentation in retinal fundus images using a deep neural network operating within the framework of instance segmentation. Considering that DNNs meant for segmentation tasks ideally require a single ground-truth for training, we developed a robust approach to fuse multiple expert annotations into a single one. We then deployed the standard MRCNN for performing simultaneous optic disc and cup detection, which was hitherto not explored. We further improved upon by the segmentation capability of the MRCNN by incorporating a residual encoder-decoder network in the mask branch. The resulting RED-RCNN architecture is more accurate than MRCNN and also more robust to variability as evidenced by the performance on a large variety of fundus image databases. The superior quality of segmentation comes from the encoder-decoder structure that is a part of the REDNet, which gives rise to a learnt, hence optimal, successive oversampling capability. The easy flow of gradients by means of skip connections is another attractive feature of the REDNet. The high performance of RED-RCNN on unseen data also emphasizes its superior generalization capability. Comparisons with state-of-the-art techniques on several publicly available databases showed that RED-RCNN is superior to several state-of-the-art methods, considering standard performance assessment measures. We also found that the superior segmentation capability of RED-RCNN translated to superior performance in two-stage glaucoma classification as well.

\begin{table}[t!]
\centering

\caption{CDR-based two-stage Glaucoma classification performance on the REFUGE dataset \cite{refuge@2018}. Se: Sensitivity; Sp:  Specificity; OCA: Overall Classification Accuracy.}
\begin{tabularx}{2.5in}{c|c|c|c}
\hline\hline
\textbf{Technique} & \textbf{Se} & \textbf{Sp} & \textbf{OCA}\\
\hline 
Expert outline & 0.849 & 0.771 & 0.779 \\
MRCNN & \textbf{0.918} & 0.769 & 0.784 \\
RED-RCNN & 0.836 & \textbf{0.926} & \textbf{0.917}  \\
\hline
\hline
\end{tabularx}
\label{tab:cdrd}
\end{table}

\section{Acknowledgments}
We would like to thank the providers of the publicly available fundus image datasets, which facilitated the experimental validation. We would also like to thank the Spectrum Lab Healthcare Team for insightful technical discussions.

\bibliographystyle{IEEEtran}
\bibliography{My_Ref2}
 
\end{document}